\documentclass[aps,prd,reprint,longbibliography,titlepage,nofootinbib]{revtex4-1}

\usepackage{amsthm}
\usepackage{amssymb}   
\usepackage{dsfont}
\usepackage{mathtools} 
\usepackage{hyperref}
\hypersetup{colorlinks=true,linkcolor=blue,urlcolor=blue,citecolor=blue}
\usepackage{accents}
\usepackage{tensor}
\usepackage{xcolor}
\usepackage{graphicx}
\usepackage[cal=boondox]{mathalfa}
\usepackage{lipsum}
\usepackage{relsize}
\usepackage{color}
\usepackage{comment}
\usepackage{nameref}
\usepackage{orcidlink}
\usepackage{appendix}

\begin{document}

%\linenumbers
\title{Hamiltonian analysis of an Effective action coupled to the Palatini action in $4$ dimensions}

\author{Ricardo Escobedo\orcidlink{0000-0001-5815-4748}}
\email[Corresponding author \\]{ricardo.escobedo@academicos.udg.mx}

\author{Roberto Santos-Silva\orcidlink{0000-0002-1265-0938}}
\email[]{roberto.santos@academicos.udg.mx}
\affiliation{Departamento de Ciencias Naturales y Exactas, CUValles,
Universidad de Guadalajara,
C.P. 46600, Ameca, Jalisco, M\'exico}

\author{Claudia Moreno\orcidlink{0000-0002-0496-032X}}
\email[]{claudia.moreno@academico.udg.mx}

\author{Rafael Hern\'andez-Jim\'enez\orcidlink{0000-0002-2740-9610}}
\email[]{rafael.hjimenez@academicos.udg.mx}

\affiliation{Departamento de F\'{i}sica, CUCEI, Universidad de Guadalajara, 44430, Guadalajara, Jalisco, M\'exico}
	
\date{\today}
	
\begin{abstract}
We carry out the canonical analysis of the coupling of a $4$-dimensional effective action that arises from a dimensional reduction of a $7$-dimensional BF theory to the Palatini action in $4$-dimensions with cosmological constant, focusing on homogeneity and isotropy in all involved fields. We employ Ostrogradsky's methodology because of the presence of second time derivatives. Through the analysis of consistency conditions, we identify three distinct and consistent possibilities. In all these scenarios, we explore the solution of second-class constraints in terms of pairs of canonical variables. Similarly, Hamiltonian analysis is performed only for the $4$-dimensional effective theory. The second-class constraints are solved in terms of canonical coordinates, revealing that this formulation aligns with one of the viable scenarios explored in the coupling to the $4$-dimensional Palatini action, entailing that the $4$-dimensional effective action nullifies gravity. 
\end{abstract}
	
\maketitle
	
%%%%%%%%%%%%%%%%%%%%%%%%%%%%%%%%%%%
\section{Introduction}\label{intro}
%%%%%%%%%%%%%%%%%%%%%%%%%%%%%%%%%%%

It is well known that Einstein's theory of general relativity can be reformulated as a modified BF theory, which was first postulated by Plebanski~\cite{10.1063/1.523215} in 1977. The gravitational field is encoded in a gauge connection and a $2$-form, where a set of constraints must be imposed to break the topology feature~\cite{Celada_2016}. In a manifestly Lorentz covariant fashion, the Hamiltonian analysis of this theory can be performed without second-class constraints~\cite{PhysRevD.101.084043} (see~\cite{Merced_Escobedo} for the case of BF gravity in $n$ dimensions). In turn, these formulations are used as a starting point for spinfoam models for quantum gravity~\cite{Perez,Rovelli_Vidotto_2014}, which are useful in the development of the path integral quantization of BF gravity and complement the canonical (or loop) approach~\cite{RovBook,ThieBook,Daniele_Oriti}. Moreover, a set of cosmological models can be developed using a generalized BF theory, in which topologically invariant rational intersection forms contribute to the cosmological constant matrices (vacuum coupling constants)~\cite{Efremov:2005ah}.

In this paper, we are interested in the study of a $4$-dimensional effective theory derived from the Kaluza-Klein compactification of a $7$-dimensional BF theory, which involves fields with second-order derivatives~\cite{efremov2014universe}. This is interesting because several advantages exist in considering a system without restrictions on the order of derivatives. These models can be used in nonlocal field theories~\cite{Erbin} and general higher-order scalar-tensor theories~\cite{PhysRevLett.114.211101,PhysRevD.89.064046,Lin_2014}, as well as infinite derivative theories in general relativity, or quantum gravity, since they are renormalizable~\cite{10.3389/fphy.2018.00077}. Nevertheless, these theories involve a ``ghost", which affects the system's physical properties~\cite{Ostrogradski}. This ``ghost" manifests itself as negative energy, linear physical momentum in the Hamiltonian, and unstable degrees of freedom. However, degenerate second-order Lagrangians generate constraints that can successfully avoid the Ostrogradsky ``ghost"~\cite{Ganz_2021}.

In this work, we explore the canonical analysis of coupling the $4$-dimensional Palatini action to a $4$-dimensional effective theory derived from a Kaluza-Klein compactification of a $7$-dimensional BF theory, assuming homogeneity and isotropy within the fields involved, taking into account the Ostrogradsky methodology due to the presence of second time derivatives in the resulting action. Analogously, we apply the same procedure only for the $4$-dimensional effective action. 

The outline of this paper is as follows. First, in Sec.~\ref{sec_II}, the resulting action of coupling the $4$-dimensional Palatini action to the effective action is derived under the assumption of homogeneity and isotropy in the fields. In Sec.~\ref{sec_III}, the canonical analysis of the resulting action is performed using Ostrogradsky and Dirac's methodology because of the presence of second-time derivatives and the singular feature of the action. As a result of the consistency conditions in the primary constraints, we find three distinct and permissible scenarios (in Appendix~\ref{non_permissible_Palatini}, the cases related to inconsistencies are explored). Then, in Sec.~\ref{sec_A_1}, we solve the second-class constraints of each consistent case using pairs of canonical coordinates. Later, in Sec.~\ref{sec_V}, the Hamiltonian analysis of the $4$-dimensional effective action is carried out following the same procedure given previously (in Appendix~\ref{non_permissible_Effremov}, the non-permissible cases are investigated). In Sec.~\ref{conlcusions}, we provide some final remarks. Moreover, to facilitate comparisons in the obtained results, the canonical analysis of the $4$-dimensional Palatini action is performed in Appendix~\ref{canonical_analysis_Palatini}, assuming homogeneity and isotropy in the fields.    

%%%%%%%%%%%%%%%%%%%%%%%%%%%%%%%%%%%%
\section{$4$-dimensional Palatini action and Effective theory}\label{sec_II}
%%%%%%%%%%%%%%%%%%%%%%%%%%%%%%%%%%%%

In this section, we perform the canonical analysis coupling gravity in the first-order formalism with or without cosmological constant to a $4$-dimensional effective action, considering homogeneity and isotropy in the fields. The latter was originally proposed in a $7$-dimensional framework given by

\begin{equation}\label{action_7}
S_7= \int_{M_7} B_3 \wedge F_4.
\end{equation}
Here $B_3$ and $F_4= dC_3$ are a $3$ and $4$ form, respectively, where $d$ is the exterior derivative. We consider an appropriate ansatz, in order to reduce via Kaluza-Klein compactification over the $3$-dimensional manifold $X$, to obtain a effective action in a $4$-dimensional manifold $M$. Let us assume that $M_7=M \times X$, and the following ansantz of the decomposition of the forms over the $X$ and $M$,
\begin{eqnarray}
    \label{B_3}B_3= \left( B -\frac{1}{2m_0 k} \star d A+d A+ \frac{1}{2m_0 k}d \star d B\right) \otimes \kappa,\\
    \label{C_3}C_3= \left( A+\frac{1}{2 m_0k} \star d B \right)\otimes \sigma,
\end{eqnarray}
where $A$ and $B$ are a $1$ and $2$ forms respectively over the $4$-dimensional manifold $M$, meanwhile $\kappa$ and $\sigma$ are closed $1$ and $2$ forms\footnote{$\kappa$ and $\sigma$ belong to the first and second De Rham cohomology groups respectively of the manifold $X$.} in the $3$-dimensional compact manifold $X$. Here $m_0$ is a topological mass parameter \cite{Diamantini} and $k$ is a coupling constant. Moreover, $\star$ is the Hodge star operator defined as:
\begin{equation}
    \star\alpha:=\frac{\sqrt{-g}}{p!(n-p)!}\alpha_{\mu_1 \cdots \mu_p}\epsilon^{\mu_1\cdots \mu_p}{}_{\nu_1\cdots\nu_{n-p}}dx^{\nu_1}\wedge\cdots\wedge dx^{\nu_{n-p}},
\end{equation}
with $\alpha$ as a $p$-form and
\begin{equation}
\epsilon^{\mu_1\cdots \mu_p}{}_{\nu_1\cdots\nu_{n-p}}=g^{\mu_1 \rho_1}\cdots g^{\nu_{\rho}\rho_{p}}\epsilon_{\rho_1 \cdots \rho_{p}\nu_1\cdots\nu_{n-p}}.
\end{equation}
Plugging back \eqref{B_3} and \eqref{C_3} into the action \eqref{action_7} we arrive at the following $4$-dimensional effective action under the assumption that $M$ has no boundary (see details in~\cite{efremov2014universe}): 

\begin{eqnarray} \label{eff_action} 
    & &S[A, B, g]=\nonumber  \\ 
    & &\int_{M}\left[ m_0 k A \wedge dB-\frac{1}{2}dA\wedge \star dA-\frac{1}{2}dB\wedge \star dB \right .\nonumber \\  & & \left .-\frac{1}{4m_0 k}\star dA \wedge d \star dB \right]. 
\end{eqnarray}
Moreover, it is well known that BF theory in four dimensions is used to model low-energy effective superconductivity \cite{Diamantini}: 
\begin{eqnarray} \label{superconductivity} 
    &&S[A, B, g]=\nonumber  \\ 
    &&\int_{M}\left[ m_0 k A \wedge dB-\frac{1}{2}dA\wedge \star dA-\frac{1}{2}dB\wedge \star dB  \right]. \nonumber \\
\end{eqnarray}

As we can observe, the actions (\ref{eff_action}) and (\ref{superconductivity}) are quite similar, differing mainly in the last term, which involves second-order derivatives. This additional term, which arises from the compactification of the theory and appears in the effective Lagrangian density, makes it necessary to introduce Ostrogradsky’s theory \cite{Ostrogradski} (see also \cite{Woodard2007} for a review) to properly define generalized momenta and perform a consistent canonical analysis, which requires pairs of canonical variables within the Hamiltonian formalism.

To carry out the canonical analysis on a $4$-dimensional manifold $M$, we label the points on it with local coordinates $(x^{\alpha})=(t:=x^0,x^a)$, where $a,b,c,\ldots=1,2,3$. In every point $x\in M$, we have the cotangent space, where we have an orthonormal frame of $1$-forms $e^I$, namely, $g=\eta_{IJ}e^I\otimes e^J$, where $g$ is the metric tensor and $(\eta_{IJ})=\mathrm{diag}(-1,1,1,1)$ is the Minkowsky metric. Thus, the indices $I,J,K,\ldots$ that take on the values $0,1,2,3$ are $SO(3,1)$-valued and are lowered and raised with $\eta_{IJ}$. Similarly, the connection $\omega^{I}{}_{J}$ is compatible with $\eta_{IJ}$, $d\eta_{IJ}-\omega^{I}{}_{I}\eta_{KJ}-\omega^{K}{}_{J}\eta_{IK}=0$, and thus $\omega_{IJ}=-\omega_{JI}$. The weight of tensor densities is either denoted by a tilde $``\sim"$ or explicitly stated elsewhere in the paper. The $SO(3,1)$ totally antisymmetric tensor $\epsilon_{IJKL}$ is such that $\epsilon_{0123}=1$. Likewise, the totally antisymmetric spacetime tensor density of weight $+1(-1)$ is denoted as $\tilde{\eta}^{\alpha\beta\gamma\delta}$ $(\underaccent{\tilde}{\eta}_{\alpha\beta\gamma\delta})$ and satisfies $\tilde{\eta}^{t123}=1$ $(\underaccent{\tilde}{\eta}_{t123}=1)$.

In the $4$-dimensional first-order formalism, general relativity with a vanishing or nonvanishing cosmological constant $\Lambda$ is described by the Palatini (or Einstein-Cartan) action:
\begin{equation}\label{Palatini_action}
    S_{P}[e,\omega]=\int_{M}\bigg[ \ast \left( e^{I}\wedge e^{J} \right)\wedge F_{IJ}-2\Lambda \rho \bigg] \,,
\end{equation}
where $F^{I}{}_{J}:=d\omega^{I}{}_{J}+\omega^{I}{}_{K}\wedge\omega^{K}{}_{J}$ is the curvature of $\omega^{I}{}_{J}$, $\rho:=(1/4!)\epsilon_{IJKL}e^I\wedge e^J \wedge e^J \wedge e^L$ is the volume form of $M$, and $``\ast"$ is the Hodge dual map\footnote{Do not confuse $\star$ with $\ast$ since the former involves a generic metric, whereas the latter is related to the Minkowski metric.} given by
\begin{equation}
    \ast(e_I\wedge e_J):=\frac{1}{2}\epsilon_{IJKL}e^K \wedge e^L,
\end{equation}
whereas the $4$-dimensional effective action is given by \eqref{eff_action} which comes from a dimensional reduction of a $7$-dimensional $BF$ theory~\cite{efremov2014universe}.

We parametrize the 16 components $e^{I}{}_{\mu}$ in terms of the tensor density $\tilde{\Pi}^{aI}$ plus the lapse and shift function $N$ and $N^a$ as
\begin{eqnarray}
    \label{e_t_I}e_{t}{}^{I} &=& Nn^{I} + N^{a}h^{1/4}h_{ab}\tilde{\Pi}^{bI}, \\
    \label{e_a_I}e_{a}{}^{I} &=& h^{1/4}h_{ab}\tilde{\Pi}^{bJ} \,,
\end{eqnarray}
where 
\begin{equation}\label{Lorentz_vector_n}
    n^I=\frac{1}{6\sqrt{h}}\epsilon_{IJKL}\underaccent{\tilde}{\eta}_{abc}\tilde{\Pi}^{aJ}\tilde{\Pi}^{bK}\tilde{\Pi}^{cL},
\end{equation}
is an internal vector that satisfies $n_I n^I=-1$ and $n_I\tilde\Pi^{aI}=0$, $h_{ab}$ of weight $-2$ is the densitized metric on the $3$-dimensional spacelike hypersurface, whose inverse is given by $\tilde{\tilde{h}}^{ab} := \tilde{\Pi}^{aI}\tilde{\Pi}^{b}{}_{I}$, and $h:=\mathrm{det}(h^{ab})$ is a tensor density of weight $4$. The maps \eqref{e_t_I} and \eqref{e_a_I} are one-to-one and invertible~\cite{PhysRevD.101.024042} (see also~\cite{PhysRevD.101.084003}). Choosing the time gauge (the time gauge is a canonical gauge, see, for instance,~\cite{HenneauxTeitelboim+1992}) by setting $n^I=\delta^I_0$, and assuming homogeneity and isotropy symmetries in the fields involved, namely, $\tilde{\Pi}^{ai}=\Pi\delta^{ai}$, with $\Pi\neq 0$, and $Q_{ai}=Q\delta_{ai}$, where $\delta_{ab}$ is the $3$-dimensional Kronocker delta, the $4$-dimensional Palatini action collapses to
\begin{equation}\label{Palatini_action_cosmological}
    S_P[\Pi,Q]=\int V_0\left[ 6\Pi\Dot{Q}-N(6\Pi^{1/2}Q^2-2\Pi^{3/2}\Lambda) \right]dt,
\end{equation}
where $V_0$ denotes the coordinate volume of the space considered, which is normalized to satisfy the condition $V_0=1$.

For the $4$-dimensional effective action \eqref{eff_action}, assuming homogeneity and isotropy in the fields, namely, $A_1=A_2=A_3=A$, $B_{12}=B_{23}=B_{31}=B$, and recalling that the relationship between the components of the metric and the orthonormal field is expressed as: 
\begin{equation}\label{relation_metric_tethrad}
    g_{\mu\nu}=e_{\mu}{}^{I}e_{\nu}{}^{J}\eta_{IJ} \,,
\end{equation}
the effective action~\eqref{eff_action} reduces to
\begin{eqnarray}\label{effective_action_cosmological}
    S_{Eff}&=&\int V_0\bigg[-6m_0 k A \Dot{B}+\frac{3}{4}\frac{\Pi^{1/2}}{N}\Dot{A}^2-\frac{\Pi^{-1/2}}{N}\Dot{B}^2\nonumber\\
    &+&\frac{\Pi^{1/2}}{4m_0 k}\Dot{A}\bigg(-\frac{1}{2N}\Pi^{-3/2}\Dot{\Pi}\Dot{B}+\frac{1}{N}\Pi^{-1/2}\Ddot{B} \nonumber\\
    &-&\frac{\Dot{N}}{N^2}\Pi^{-1/2}\Dot{B}\bigg)\bigg] dt,
\end{eqnarray}
with $V_0=1$ as the normalized volume of the space considered. 

In this paper, we focus on the coupling of the effective action to the $4$-dimensional Palatini action, considering homogeneity and isotropy in all fields involved. Consequently, the theory we will study is given by the action principle
\begin{eqnarray}\label{action1}
    && \hspace{-0.5cm} S_{\mathrm{P}+Eff} = \int_{M}\bigg[ 6\Pi \Dot{Q}-6m_0 kA\Dot{B}+\frac{3}{4}\frac{\Pi^{1/2}}{N}\Dot{A}^2-\frac{\Pi^{-1/2}}{N}\Dot{B}^2 \nonumber \\
  && \hspace{-0.2cm} +\frac{\Pi^{1/2}}{4m_0 k N}\Dot{A} \bigg( -\frac{1}{2N}\Pi^{-3/2}\Dot{\Pi}\Dot{B}+\frac{1}{N}\Pi^{-1/2}\Ddot{B} \nonumber \\ 
  && \hspace{-0.2cm} - \frac{\Dot{N}}{N^2}\Pi^{-1/2}\Dot{B} \bigg)
    - 2N\Pi^{1/2}\left( 3Q^2-\Pi \Lambda \right) \bigg]dt \,.\nonumber  \\
\end{eqnarray}

\section{Canonical analysis}\label{sec_III}
The canonical formulation of the higher derivative action \eqref{action1} can be obtained directly following the method of Ostrogradsky~\cite{Ostrogradski} (see, for instance~\cite{Woodard2007} for a review). In this formalism, there are $20$ canonical variables, which $10$ of them are the set of coordinates $Q, q:=\Dot{Q},\Pi, p:=\Dot{\Pi}, A, \alpha:=\Dot{A}, B, \beta:= \Dot{B}, N,n:=\Dot{N}$, and the other $10$ are their corresponding canonical momenta given as follows:

\begin{eqnarray}
    \label{Q_Momemtum}&&P_{Q}=\frac{\delta L}{\delta \Dot{Q}}=\frac{\partial L}{\partial \Dot{Q}}=6\Pi \,, \\
    \label{q_momentum}&& P_q=\frac{\delta L}{\delta \Ddot{Q}}=\frac{\partial L}{\partial \Ddot{Q}}=0 \,,\\
    \label{Pi_momemtum}&& P_{\Pi}=\frac{\delta L}{\delta \Dot{\Pi}}=\frac{\partial L}{\partial \Dot{\Pi}}=-\frac{1}{8m_0 k}\frac{\Dot{A}\Dot{B}}{N^2 \Pi} \,,\\
    \label{p_momemtum}&& P_{p}=\frac{\delta L}{\delta \Ddot{\Pi}}=\frac{\partial L}{\partial \Ddot{\Pi}}=0 \,, \\   
    \label{A_momemtum}&& P_A= \frac{\delta L}{\delta \Dot{A}}=\frac{\partial L}{\partial \Dot{A}}=\frac{3}{2}\frac{\Pi^{1/2}}{N}\Dot{A}\nonumber \\ &&+\frac{\Pi^{1/2}}{4m_0 kN}\bigg( -\frac{1}{2N}\Pi^{-3/2}\Dot{\Pi}\Dot{B}+\frac{1}{N}\Pi^{-1/2}\Ddot{B}\nonumber \\ && -\frac{\Dot{N}}{N^2}\Pi^{-1/2}\Dot{B} \bigg),\\ \label{alpha_momemtum}&& P_{\alpha}=\frac{\delta L}{\delta \Ddot{A}}=\frac{\partial L}{\partial \Ddot{A}}=0 \,, \\
    \label{B_momemtum}&& P_B=\frac{\delta L}{\delta \Dot{B}}=\frac{\partial L}{\partial \Dot{B}}-\frac{d}{dt}\left( \frac{\partial L}{\partial \Ddot{B}} \right)=\nonumber \\&& -6m_0 A-\frac{2}{N}\Pi^{-1/2}\Dot{B}-\frac{\Dot{A}\Dot{\Pi}}{8m_0 kN^2}+\frac{1}{4m_0 k}\frac{\Dot{A}\Dot{N}}{N^3} \nonumber \\&& -\frac{1}{4m_0 k}\frac{\Ddot{A}}{N^2} \,, \\
    \label{beta_momemtum}&& P_{\beta}=\frac{\delta L}{\delta \Ddot{B}}=\frac{\partial L}{\partial \Ddot{B}}=\frac{1}{4m_0 k}\frac{\Dot{A}}{N^2} \,,\\
    \label{N_momemtum}&& P_{N}=\frac{\delta L}{\delta \Dot{N}}=\frac{\partial L}{\partial \Dot{N}}=-\frac{1}{4m_0 k}\frac{\Dot{A}\Dot{B}}{N^3} \,, \\
    \label{n_momemtum}&& P_n=\frac{\delta L}{\delta \Ddot{N}}=\frac{\partial L}{\partial \Ddot{N}}=0 \,.
\end{eqnarray}
More explicitly, the set of canonical variables with their respective canonical momentum computed in \eqref{Q_Momemtum}-\eqref{n_momemtum} are given by the $10$ following pairs:
$(Q, P_Q)$, $(q,P_q)$, $(\Pi, P_{\Pi})$, $(p,P_p)$, $(A, P_A)$, $(\alpha, P_{\alpha})$, $(B, P_B)$, $(\beta, P_{\beta})$, $(N, P_N)$ and $(n,P_n)$.
The set of primary constraints is derived from the definition of the canonical momentum given by \eqref{Q_Momemtum}-\eqref{p_momemtum}, and \eqref{beta_momemtum}-\eqref{n_momemtum},
reading:
\begin{eqnarray}
    \label{1_prim_comstraint}&&\phi_1:=P_Q-6\Pi \,, \\
    &&\phi_2:=P_q \,, \\
    &&\phi_3:=P_{\Pi}+\frac{1}{8m_0 k}\frac{\alpha \beta}{N^2 \Pi} \,, \\
    &&\phi_4:= P_p \,, \\
    &&\phi_5:=P_{\alpha} \,, \\
    \label{phi_6_Palatini_Effremov}&&\phi_6:=P_{\beta}-\frac{1}{4m_0 k}\frac{\alpha}{N^2} \,, \\
    &&\phi_7:=P_N+\frac{1}{4m_0 k}\frac{\alpha \beta}{N^3} \,, \\
    \label{8_prim_constraint}&& \phi_8:= P_n \,.
\end{eqnarray}
Using the definition provided in $P_A$, one can derive $\Ddot{B}=\Dot{\beta}$ in terms of the coordinates of the phase space, resulting in
\begin{eqnarray}
    \label{dot_beta}\Dot{\beta}=4m_0 kN^2 P_A-6m_0 k \Pi^{1/2}N\alpha+\frac{1}{2}\frac{p\beta}{\Pi}+\frac{n\beta}{N} \,. \nonumber \\
\end{eqnarray}
Analogously to the previous equation, we derive $\Ddot{A}=\Dot{\alpha}$ yielding,
\begin{eqnarray}\label{dot_alpha}
    \Dot{\alpha} &=& -4m_0 kN^2 P_B-24m_0^2 k^2N^2 A \nonumber \\ && -8m_0 kN\Pi^{-1/2}\beta -\frac{1}{2}\alpha p+\frac{\alpha n}{N} \,.
\end{eqnarray}
Given that there are $8$ primary constraints, we proceed following Dirac's standard procedure~\cite{dirac1964lectures,HenneauxTeitelboim+1992}. 
The Hamiltonian of Ostrogradsky is established through Legendre transformation as follows:
\begin{eqnarray}  H_{\mathrm{Ost}}&=&p_{i}\Dot{q}^{i}-L \nonumber\\
    &=&P_Q q+P_q\Dot{q}+P_{\Pi}p+P_p\Dot{p}-4m_0kP_{\alpha}P_BN^2  \nonumber\\
    &-&24m_0^2kN^2AP_{\alpha}-8m_0kN\Pi^{-1/2}\beta P_{\alpha}-\frac{1}{2}\alpha\beta P_{\alpha} \nonumber\\
    &+&\frac{\alpha n}{N}P_{\alpha} +P_{B}\beta +4m_0 kN^2 P_{\beta}P_{A} \nonumber\\
    &-&6m_0k\Pi^{1/2}\alpha N P_{\beta}+\frac{1}{2}\frac{p\beta}{\Pi}P_{\beta}+\frac{n\beta}{N}P_{\beta} \nonumber\\
   &+&P_{N}n+P_n \Dot{n} -6\Pi q+6m_0 kA\beta+\frac{3}{4}\frac{\Pi^{1/2}}{N}\alpha^2 \nonumber\\
   &+&\frac{\Pi^{-1/2}}{N}\beta^2 + \frac{\alpha}{4m_0 k}\frac{n\beta}{N^3} \nonumber \\&+&\left( 6N\Pi^{1/2}Q^2-2N\Pi^{3/2}\Lambda \right),
\end{eqnarray}
where \eqref{dot_beta} and \eqref{dot_alpha} were used. Thus, by employing the primary constraints \eqref{1_prim_comstraint}-\eqref{8_prim_constraint}, the primary Hamiltonian acquires the following form,
\begin{eqnarray}\label{primary_hamiltonian}
    H_{P}&=&P_Q q+P_{\Pi}p+P_{B}\beta +4m_0 kN^2 P_{\beta}P_{A}\nonumber\\
    &-&6m_0k\Pi^{1/2}\alpha N P_{\beta}+\frac{1}{2}\frac{p\beta}{\Pi}P_{\beta}+\frac{n\beta}{N}P_{\beta} \nonumber\\
    &+& P_{N}n - 6\Pi q +6m_0 kA\beta+\frac{3}{4}\frac{\Pi^{1/2}}{N}\alpha^2 \nonumber\\
    &+&\frac{\Pi^{-1/2}}{N}\beta^2 +\frac{\alpha}{4m_0 k}\frac{n\beta}{N^3} \nonumber \\
    &+&\left( 6N\Pi^{1/2}Q^2-2N\Pi^{3/2}\Lambda \right) \nonumber\\&+&\lambda_1\phi_1+\lambda_2\phi_2+\lambda_3\phi_3+\lambda_4\phi_4+\lambda_5\phi_5+\lambda_6\phi_6 \nonumber\\
    &+&\lambda_7\phi_7+\lambda_8\phi_8,
\end{eqnarray} 
where $\lambda'$s represent the Lagrange multipliers. To maintain consistency, all constraints must be preserved weakly on time, leading to:

\begin{eqnarray}
\label{dot_phi_1_first}\Dot{\phi}_1 &=& -6(2N\Pi^{1/2}Q+p)-6\lambda_3, \\
    \label{dot_phi_2_first}\Dot{\phi}_2&=& -\phi_1\approx 0, \\
    \label{dot_phi_3_first}\Dot{\phi}_3 &=& \left(3m_0 k \Pi^{-1/2}\alpha N+\frac{1}{2}\frac{p\beta}{\Pi^2}\right)P_{\beta} \nonumber\\
    &+&\frac{1}{2}\frac{\Pi^{-3/2}}{N}\beta^2 -\frac{9}{8}\frac{\alpha^2}{\Pi^{1/2}N}-3N\Pi^{-1/2}Q^2 \nonumber \\ &+& 3N \Pi^{1/2}\Lambda  + \frac{1}{2}\frac{\alpha P_A}{\Pi} \nonumber\\
    &-&\frac{1}{8m_0 k}\frac{\alpha \beta}{\Pi N^2}\left( \frac{n}{N}+\frac{p}{2\Pi} \right)+6\lambda_1 \nonumber \\ &+& \frac{1}{8m_0 k}\frac{\beta}{N^2 \Pi}\lambda_5
    + \frac{1}{8m_0 k}\frac{\alpha}{N^2 \Pi}\lambda_6  \nonumber\\ &-& \frac{1}{4m_0 k}\frac{\alpha \beta}{\Pi N^3}\lambda_{7}, \\
    \label{dot_phi_4_first}\Dot{\phi}_4 &=&-\phi_3-\frac{\beta}{2\Pi}\phi_6\approx 0, \\
\label{dot_phi_5_first}\Dot{\phi}_{5}&=& -\frac{1}{4m_0 k}\frac{1}{N^2}\left[ \frac{\beta}{N}(n-\lambda_7)+\frac{\beta}{2}\lambda_3 -\lambda_6 \right],\nonumber \\   \label{dot_phi_6_first}\Dot{\phi}_6&=&-P_B-\frac{1}{2}\frac{pP_{\beta}}{\Pi}-\frac{n}{N}P_{\beta}-6m_0 kA\nonumber\\
    &-&\frac{2\Pi^{-1/2}\beta}{N}+\frac{1}{4m_0 k}\frac{\alpha n}{N^3} \nonumber \\ &-& \frac{1}{4m_0 k}\frac{\alpha}{N^2}\left( \frac{\lambda_3}{2\Pi}-\frac{\lambda_7}{N} \right) 
    -\frac{1}{4m_0 k}\frac{1}{N^2}\lambda_5,
    \end{eqnarray}
    
    \begin{eqnarray}
    \label{dot_phi_7_first}\label{dot_phi_7}\Dot{\phi}_7 &=& \bigg( -8m_0 kNP_A+6m_0k\Pi^{1/2}\alpha+\frac{n\beta}{N^2} \bigg)P_{\beta}\nonumber\\
    &+&\frac{\Pi^{-1/2}}{N^2}\beta^{2}-\frac{3}{4}\frac{\Pi^{1/2}}{N^2}\alpha^2 \nonumber \\ &+& \frac{1}{4m_0 k}\frac{\alpha \beta}{N^3}\bigg( \frac{n}{N}+\frac{p}{2\Pi} \bigg) 
    - 6\Pi^{1/2}Q^2 \nonumber \\ &+& 2\Pi^{3/2}\Lambda  + \frac{\alpha}{N}P_A+\frac{1}{4m_0 k}\frac{\alpha\beta}{N^3\Pi}\lambda_3\nonumber\\
    &+&\frac{1}{4m_0 k}\frac{\beta}{N^3}\lambda_5-\frac{1}{4m_0 k}\frac{\alpha}{N^3}\lambda_6,\\
\label{dot_phi_8_first}\label{dot_phi_8} \Dot{\phi}_8&=&-\frac{1}{4m_0 k}\frac{\alpha\beta}{N^3}-\left(\phi_7+\frac{\beta}{N}\phi_6\right).
\end{eqnarray}
The consistency relation for $\phi_1$ implies the following expression, 
\begin{equation}\label{lambda_3.1}
    \lambda_3=-(2N\Pi^{1/2}Q+p).
\end{equation}
Notice that $\Dot{\phi}_8$ given by \eqref{dot_phi_8_first} does not impose conditions on any of the Lagrange multipliers. As a result, its consistency relation leads to a secondary constraint. Given that $N\neq 0$, the secondary constraint is reduced to $\alpha\beta \approx0$, giving rise to three cases to consider:\\

i) A secondary constraint arises when $\alpha\approx 0$, leaving $\beta\neq 0$ as a free variable.\\

ii) A secondary constraint occurs when $\beta\approx 0$, which allows $\alpha\neq 0$ to remain a free variable.\\

iii) Two secondary constraints emerge, given by $\alpha\approx0$ and $\beta\approx0$.\\

It turns out that case (ii) is not a permissible case to perform (see Appendix~\ref{non_permissible_Palatini}). The permissible cases will be discussed below.
\subsection{Case $\alpha \approx 0$ and $\beta\neq 0$.}
The consistency relation for this secondary constraint $\psi:=\alpha$ collapses to:
\begin{equation}
    \Dot{\psi}=\lambda_5,
\end{equation}
implying that
\begin{equation}\label{lambda_5.1}
    \lambda_5=0.
\end{equation}
Considering the secondary constraint $\psi=\alpha$ and the result $\lambda_5=0$, the previous consistency relations \eqref{dot_phi_1_first}-\eqref{dot_phi_8_first} can be expressed as follows:

\begin{eqnarray}
    \Dot{\phi}_1
    &\approx& -6(2N\Pi^{1/2}Q+p)-6\lambda_3, \\
    \Dot{\phi}_2&\approx& 0, \\
    \Dot{\phi}_3 &\approx& \frac{1}{2}\frac{\Pi^{-3/2}}{N}\beta^2-3N\Pi^{-1/2}Q^2\nonumber\\
    &+&3N\Pi^{1/2}\Lambda +6\lambda_1,\\
    \Dot{\phi}_4 &\approx& 0, \\
    \label{Consistency_phi_5}\Dot{\phi}_{5}&\approx& -\frac{1}{4m_0 k}\frac{1}{N^2}\left[ \frac{\beta}{N}(n-\lambda_7)+\frac{\beta}{2}\lambda_3 -\lambda_6 \right],\nonumber \\   \\
    \label{evolution_6}\Dot{\phi}_6&\approx&-P_B-6m_0 kA-\frac{2\Pi^{-1/2}\beta}{N},\\
    \label{dot_phi_7}\Dot{\phi}_7 &\approx& \frac{\Pi^{-1/2}}{N^2}\beta^{2}-6\Pi^{1/2}Q^2+2\Pi^{3/2}\Lambda, \\
   \label{dot_phi_8} \Dot{\phi}_8&\approx&0,\\
   \Dot{\psi}&=&
\lambda_5=0.
\end{eqnarray}
At this stage of the analysis, we can solve for the Lagrange multipliers $\lambda_3$ and $\lambda_1$ from $\phi_1$ and $\phi_3$ respectively, yielding:
\begin{eqnarray}
    \label{sol_lambda_3}\lambda_3&=& -(2N \Pi^{1/2} Q+p), \\
    \label{sol_lambda_1}\lambda_1&=&-\frac{1}{12}\frac{\Pi^{-3/2}}{N}\beta^2+\frac{1}{2}N\Pi^{-1/2}Q^2\nonumber\\
    &-&\frac{1}{2}N\Pi^{1/2}\Lambda.
\end{eqnarray}
Since $\Dot{\phi}_6$ and $\dot{\phi}_7$ do not impose conditions on the involved Lagrange multipliers, they create new tertiary constraints given by:
\begin{eqnarray}
    \label{tertiarity_1}\chi_1&:=& P_B+6m_0 k A+\frac{2\Pi^{-1/2}}{N}\beta, \\
    \label{tertiary_2}\chi_2&:=& \frac{\Pi^{-1/2}}{N^2}\beta^2 -6\Pi^{1/2}Q^2+2\Pi^{3/2}\Lambda.
\end{eqnarray}
It is important to emphasize that \eqref{sol_lambda_1} and \eqref{tertiary_2} are related to each other as follows:
\begin{eqnarray}\label{lambda_1.1}
    \lambda_1&=&-\frac{N}{12\Pi}\chi_2-\frac{1}{3}N\Pi^{1/2}\Lambda \nonumber\\
    &\approx& -\frac{1}{3}N\Pi^{1/2}\Lambda.
\end{eqnarray}
Using the solutions for $\lambda_1$ and $\lambda_3$ given by \eqref{sol_lambda_3} and \eqref{sol_lambda_1} respectively,
the temporal evolution of the tertiary constraints $\chi_1$ and $\chi_2$ leads to the following expressions:
\begin{eqnarray}
\label{dot_chi_1_beta_neq_0}\Dot{\chi_1}&=&\frac{2\Pi^{-1/2}}{N}\bigg( 4m_0 kN^2 P_A+\frac{1}{2}\frac{p\beta}{\Pi}\nonumber\\
&+&\frac{n\beta}{N}+\lambda_6 \bigg)+\frac{2\beta Q}{\Pi}-\frac{2\Pi^{-1/2}\beta}{N^2}\bigg( n+\lambda_7 \bigg), \\
\label{consistency_chi_2}\Dot{\chi_2}&=& -12\Pi^{1/2}Qq.
\end{eqnarray}
Notice that the consistency relation $\Dot{\chi}_1$ includes conditions for the Lagrange multipliers $\lambda_6$ and $\lambda_7$. With $\lambda_3$ already calculated which solution is \eqref{sol_lambda_3}, we can determine $\lambda_6$ and $\lambda_7$ using \(\Dot{\phi_5}\) from \eqref{Consistency_phi_5} and \(\Dot{\chi_1}\) from \eqref{dot_chi_1_beta_neq_0}, giving as a result:
\begin{eqnarray}
    \label{lambda_6.1}\lambda_6&=&\frac{\beta}{2}\left( \frac{n}{N}-\frac{p}{\Pi} \right)-2m_0 kN^2 P_A \nonumber\\
    &-&\beta N\Pi^{-1/2}Q, \\
    \label{sol_lambda_7}\lambda_7&=&\frac{2}{\beta}m_0 k N^3 P_A +\frac{1}{2}n.
\end{eqnarray}
 At this stage of the procedure, it is notorious that \eqref{consistency_chi_2} leads to quaternary constraints. Nonetheless, since $\Pi\neq 0$, three distinct scenarios can be systematically analyzed:\\

i) A quaternary constraint occurs when $Q\approx 0$, leaving $q\neq 0$ as a free variable.\\

ii) A quaternary constraint emerges when $q\approx0$,which allows $Q\neq0$ to be a free variable.\\

iii) Two quaternary constraints arise, given by $Q\approx0$ and $q\approx0$.\\

It turns out that only the third case is inconsistent (see Appendix~\ref{non_permissible_Palatini}). The first and second cases are consistent, and the development of such cases will be performed below.  
\subsubsection{Case $Q\approx0$ and $q\neq 0$.}\label{Case_1_permissible}
Here we only have one quaternary constraint $\xi$ given by:
\begin{equation}
    \xi:=Q.
\end{equation}
Thus, applying the consistency relation to such a constraint, we arrive at the expression
\begin{equation}
    \Dot{\xi}=q-\frac{1}{3}N\Pi^{1/2}\Lambda,
\end{equation}
where the explicit value of $\lambda_1$ given in \eqref{lambda_1.1} was used. At this stage, we can conclude that $\Lambda\neq0$. Otherwise, $q$ would be a constraint. Thus, since the latter expression does not impose any conditions on the Lagrange multipliers, a new constraint arises, which we denote by $\Omega$, namely:
\begin{equation}
    \Omega:=q-\frac{1}{3}N\Pi^{1/2}\Lambda.
\end{equation}
The consistency relation of the latter displays the following:
\begin{equation}
    \Dot{\Omega}=\lambda_2-\frac{1}{3}\Pi^{1/2}\Lambda(\frac{3}{2}n+\frac{2}{\beta}m_0 kN^3 P_A).
\end{equation}
Solving the latter relation for $\lambda_2$ yields,
\begin{equation}\label{solution_lambda_2_1}
    \lambda_2=\frac{1}{3}\Pi^{1/2}\Lambda(\frac{3}{2}n+\frac{2}{\beta}m_0kN^3P_A).
\end{equation}
Furthermore, from \eqref{sol_lambda_3} and \eqref{lambda_6.1}, the solutions for $\lambda_3$ and $\lambda_6$ collapse to:
\begin{eqnarray}
\label{solution_lambda_3_Q_van} \lambda_3&=&-p, \\
\label{solution_lambda_6_Q_van}\lambda_6&=&\frac{\beta}{2}\bigg( \frac{n}{N}-\frac{p}{\Pi} \bigg)-2m_0kN^2P_A.
\end{eqnarray}
The procedure for finding constraints using the consistency relation concludes here, since all the constraints involved are preserved weakly in time, and there are no further conditions for the Lagrange multipliers.

In this scenario, we have $\Lambda\neq0$, and the complete set of constraints can be rewritten as follows:
\begin{eqnarray}
    \label{constraint_1_Q_van}\phi_1&=&P_Q-6\Pi, \\
    \phi_2&=& P_q, \\
    \phi_3&=&P_{\Pi},\\
    \phi_4&=& P_p, \\ 
    \phi_5&=&P_{\alpha}, \\
    \phi_6&=&P_{\beta}, \\
    \phi_7&=&P_N, \\
    \phi_8&=&P_n, \\
    \psi&=&\alpha, \\
\chi_1&=&P_B+6m_0kA+\frac{2\Pi^{-1/2}}{N}\beta, \\
    \chi_2&=&\frac{\Pi^{-1/2}}{N^2}\beta^2+2\Pi^{3/2}\Lambda, \\
    \xi&=&Q, \\
    \label{constraint_13_Q_van}\Omega&=&q-\frac{1}{3}N\Pi^{1/2}\Lambda.
\end{eqnarray}

The constraints $\phi_3$, $\phi_6$ and $\phi_7$ were redefined by considering the presence of $\psi=\alpha$. In what follows, to find a complete set of first-class and second-class constraints, we compute the Poisson brackets among the constraints \eqref{constraint_1_Q_van}-\eqref{constraint_13_Q_van}, leading to

\begin{widetext}
\begin{equation}\label{matrix_form_2.1}
\scriptsize{
[C_A,C_B]=
\begin{pmatrix}
0 & 0 & -6&0 &0&0&0&0&0&0&0&-1&0\\
0 & 0 & 0&0&0&0&0&0&0&0&0&0&-1\\
6&0&0& 0&0&0&0&0&0&\frac{\beta}{N}\Pi^{-3/2}&\frac{1}{2}\frac{\beta^2}{N^2}\Pi^{-3/2}-3\Pi\Lambda&0&\frac{1}{6}N\Pi^{-1/2}\Lambda \\
0&0&0&0&0&0&0&0&0&0&0&0&0 \\
0&0&0&0&0&0&0&0&-1&0&0&0&0 \\
0&0&0&0&0&0&0&0&0&-\frac{2\Pi^{-1/2}}{N}&-\frac{2\Pi^{-1/2}}{N^2}\beta&0&0 \\
0&0&0&0&0&0&0&0&0&\frac{2\Pi^{-1/2}}{N^2}\beta&\frac{2\Pi^{-1/2}}{N^3}\beta^2&0&\frac{1}{3}\Pi^{1/2}\Lambda \\
0&0&0&0&0&0&0&0&0&0&0&0&0 \\
0&0&0&0&1&0&0&0&0&0&0&0&0 \\
0&0&-\frac{\beta}{N}\Pi^{-3/2}&0&0&\frac{2\Pi^{-1/2}}{N}&-\frac{2\Pi^{-1/2}}{N^2}\beta&0&0&0&0&0&0 \\
0&0&-\frac{1}{2}\frac{\beta^2}{N^2}\Pi^{-3/2}+3\Pi\Lambda&0&0&\frac{2\Pi^{-1/2}}{N^2}\beta&-\frac{2\Pi^{-1/2}}{N^3}\beta^2&0&0&0&0&0&0\\
1&0&0&0&0&0&0&0&0&0&0&0&0\\
0&1&-\frac{1}{6}N\Pi^{-1/2}\Lambda&0&0&0&-\frac{1}{3}\Pi^{1/2}\Lambda&0&0&0&0&0&0\\
\end{pmatrix} \,,
}
\end{equation}
\end{widetext}
where the indices $A$ and $B$ run from $1$ to $13$. Each number denotes the constraints $\phi_1$, $\phi_2$, $\phi_3$, $\phi_4$, $\phi_5$, $\phi_6$, $\phi_7$, $\phi_8$, $\psi$, $\chi_1$, $\chi_2$, $\xi$ and $\Omega$ respectively. The rank is $10$, with nullity $13-10=3$, which implies that the eigenvalue of three independent eigenvectors vanishes. 
As a result, there are only three independent combinations of constraints that are first-class, namely
\begin{eqnarray}
    C_1^{FC}&:=&\phi_4=P_p, \\
    C_2^{FC}&:=&\phi_8=P_n, \\
    C_3^{FC}&:=&\frac{1}{3}\Pi^{1/2}\Lambda\phi_2+\frac{\beta}{N}\phi_6+\phi_7 \nonumber\\
    &=&\frac{1}{3}\Pi^{1/2}P_q\Lambda+\frac{\beta}{N}P_{\beta}+P_N,
\end{eqnarray}
where the superscript $FC$ stands for first-class. 
Consequently, there are ten remaining linear combinations of constraints, which we take to be
\begin{eqnarray}
\label{C1SC_Q_van} C_1^{SC}&:=&\phi_1=P_Q-6\Pi, \\
\label{C2SC_Q_van}C_2^{SC}&:=&\phi_2=P_q, \\
\label{C3SC_Q_van}C_3^{SC}&:=&\phi_3=p_{\Pi}, \\
\label{C4SC_Q_van}C_4^{SC}&:=&\phi_5=P_{\alpha}, \\
\label{C5SC_Q_van}C_5^{SC}&:=&\phi_7=P_N, \\
\label{C6SC_Q_van}C_6^{SC}&=&\psi=\alpha, \\
\label{C7SC_Q_van}C_7^{SC}&=&\chi_1=P_B+6m_0kA+\frac{2\Pi^{-1/2}}{N}\beta, \\
\label{C8SC_Q_van}C_8^{SC}&:=&\chi_2=\frac{\Pi^{-1/2}}{N^2}\beta^2+2\Pi^{3/2}\Lambda, \\
\label{C9SC_Q_van}C_9^{SC}&:=&\xi=Q, \\
\label{C10SC_Q_van}C_{10}^{SC}&:=&q-\frac{1}{3}N\Pi^{1/2}\Lambda,
\end{eqnarray}
where the superscript $SC$ stands for second-class.
The total Hamiltonian is obtained by substituting the above conditions of the Lagrange multipliers $\lambda_1$, $\lambda_2$, $\lambda_3$, $\lambda_5$, $\lambda_6$ and $\lambda_7$ given by \eqref{lambda_1.1}, \eqref{solution_lambda_2_1}, \eqref{solution_lambda_3_Q_van}, \eqref{lambda_5.1}, \eqref{solution_lambda_6_Q_van} and \eqref{sol_lambda_7} respectively into the primary Hamiltonian \eqref{primary_hamiltonian} leading to: 
\begin{equation}
H_T=H'+\lambda_4C_1^{FC}+\lambda_8C_2^{FC},
\end{equation}
with
\begin{widetext}
\begin{eqnarray}
    H'&=& \bigg( C_3^{FC}-\frac{1}{3}\Pi^{1/2}\Lambda C_2^{SC}-C_3^{SC} \bigg)\bigg[ \frac{2m_0 kN^3}{\beta}\bigg( P_A-3\Pi^{1/2}C_6^{SC} \bigg)+\frac{3}{2}n \bigg]+C_1^{SC}C_{10}^{SC}+\beta C_7^{SC} \nonumber\\
    &-&NC_8^{SC}+nC_5^{SC}+\frac{3}{4}\frac{\Pi^{1/2}}{N}(C_6^{SC})^2+\frac{1}{4m_0k}\frac{n\beta}{N^3}C_6^{SC}+6N\Pi^{1/2}(C_9^{SC})^2 \nonumber\\
    &+&\frac{1}{3}\Pi^{1/2}\Lambda
\bigg( \frac{3}{2}n+\frac{2}{\beta}m_0kN^3P_A \bigg)C_2^{SC}-\frac{p\beta}{8m_0k\Pi}C_6^{SC} -\frac{1}{4m_0k}\frac{\psi}{N^2}\bigg[ \frac{\beta}{2}\bigg( \frac{n}{N}-\frac{p}{\Pi} \bigg)-2m_0kN^2P_A \bigg] \nonumber\\
&+&\bigg( \frac{2}{\beta}m_0kN^3P_A+\frac{1}{2}n \bigg) \bigg( C_5^{SC}+\frac{\beta}{4m_0 kN^3}C_6^{SC} \bigg),
\end{eqnarray}    
\end{widetext}
as the first-class Hamiltonian. It is worthwhile that this formulation is a gauge theory, since the first-class Hamiltonian $H'$ is formed by constraints only.

\subsubsection{Case  $Q\neq 0$ and $q\approx 0$.}\label{Case_2_permissible}
In this case, the quaternary constraint denoted by $\xi$ is given by:
\begin{equation}
    \xi:=q.
\end{equation}
Their corresponding consistency relation yields
\begin{equation}
    \Dot{\xi}=\lambda_2,
\end{equation}
which gives the following condition:
\begin{equation}\label{lambda_2.1}
    \lambda_2=0.
\end{equation}
The procedure ends here since all the constraints involved here are weakly preserved in time. The system is consistent, and the analysis can proceed. It is also important to note that $\Lambda$ can take any value. Bearing in mind that $\psi = \alpha$, the set of constraints can be rewritten as follows:
\begin{eqnarray}
    \label{constraint_1_q_van}\phi_1&=&P_Q-6\Pi, \\
    \phi_2&=& P_q, \\
    \phi_3&=& P_{\Pi}, \\
    \phi_4&=& P_p, \\
    \phi_5&=& P_{\alpha}, \\
    \phi_6&=& P_{\beta}, \\
    \phi_7&=& P_N, \\
    \phi_8&=& P_n, \\
    \psi&=& \alpha, \\
    \chi_1&=& P_B+6m_0 kA +\frac{2\Pi^{-1/2}}{N}\beta, \\
    \chi_2&=& \frac{\Pi^{-1/2}}{N^2}\beta^2-6\Pi^{1/2}Q^2+2\Pi^{3/2}\Lambda, \\
    \label{constraint_12_q_van}\xi&=& q.
\end{eqnarray}
To find a complete set of first-class and second-class constraints, we compute the Poisson bracket among the constraints \eqref{constraint_1_q_van}-\eqref{constraint_12_q_van}, yielding: 

\begin{widetext}

\begin{equation}\label{matrix_form}
\scriptsize{
[C_A,C_B]=
\begin{pmatrix}
0 & 0 & -6&0 &0&0&0&-1&0&0&12\Pi^{1/2}Q&0\\
0 & 0 & 0&0&0&0&0&0&0&0&0&-1\\
6 & 0 & 0&0&0&0&0&0&0&\frac{\Pi^{-3/2}}{N}\beta&\frac{1}{2}\frac{\Pi^{-3/2}}{N^2}\beta^2-3\Pi^{1/2}\Lambda&0 \\
0&0&0& 0&0&0&0&0&0&0&0&0 \\
0&0&0&0&0&0&0&0&-1&0&0&0 \\
0&0&0&0&0&0&0&0&0&-\frac{2\Pi^{-1/2}}{N}&-\frac{2\Pi^{-1/2}\beta}{N^2}&0 \\
0&0&0&0&1&0&0&0&0&\frac{2\Pi^{-1/2}\beta}{N^2}&\frac{2\Pi^{-1/2}}{N^3}\beta^2&0 \\
0&0&0&0&0&0&0&0&0&0&0&0 \\
0&0&0&0&1&0&0&0&0&0&0&0 \\
0&0&-\frac{\Pi^{-3/2}}{N}\beta&0&0&\frac{2\Pi^{-1/2}}{N}&-\frac{2\Pi^{-1/2}}{N^2}\beta^2&0&0&0&0&0\\
-12\Pi^{1/2}Q&0&-\frac{1}{2}\frac{\Pi^{-3/2}}{N^2}\beta^2+3\Pi^{1/2}\Lambda&0&0&\frac{2\Pi^{-1/2}}{N^2}\beta&-\frac{2\Pi^{-1/2}}{N^3}\beta^2&0&0&0&0&0\\
0&1&0&0&0&0&0&0&0&0&0&0\\
\end{pmatrix} \,,
}
\end{equation}
\end{widetext}
where the indices $A$ and $B$ run from $1$ to $12$. Each number represents the constraints $\phi_1$, $\phi_2$, $\phi_3$, $\phi_4$, $\phi_5$, $\phi_6$, $\phi_7$, $\phi_8$, $\psi$, $\chi_1$, $\chi_2$ and $\xi$ respectively. The rank is $8$, with nullity $12-8=4$, which implies that the eigenvalue of three independent eigenvectors vanishes. Due to this, there are four independent combinations of constraints that are first-class, namely
\begin{eqnarray}
   C_{1}^{FC}&=&(P_Q-6\Pi)(\beta^2 +6\Pi^2 N^2 \Lambda)+Q(2\Pi^{1/2}P_{\Pi} \nonumber \\ &+&P_{\beta}\beta\Pi^{-1/2}-6\Pi^{1/2}Q)  +2\Pi^{3/2}\Lambda \nonumber\\
   &&-\frac{\beta}{N}(P_B +6m_0 kA+\frac{\Pi^{-1/2}\beta}{N}) , \\
   C_2^{FC}&=&P_{\Pi}, \\
   C_3^{FC}&=&P_p\frac{\beta}{N}+P_N, \\
   C_4^{FC}&=&P_p,
\end{eqnarray}
where the superscript $FC$ stands for first-class. Therefore, there are $8$ second-class constraints, which we take to be
\begin{eqnarray}
    \label{SC1.1}C_1^{SC}&=&P_q, \\
    \label{SC2-1}C_2^{SC}&=&P_{\alpha},\\ \label{SC3.1}C_3^{SC}&=&\alpha,\\
    \label{SC4.1}C_4^{SC}&=&q,\\
    \label{SC5.1}C_5^{SC}&=&P_{\beta}, \\  \label{SC6.1}C_6^{SC}&=&\frac{\Pi^{-1/2}}{N^2}\beta^2-6\Pi^{1/2}Q^2+2\Pi^{3/2}\Lambda, \\
     \label{SC7.1}C_7^{SC}&=&P_B+6m_0 k A+\frac{2\Pi^{-1/2}}{N}\beta,\\
    \label{SC8.1}C_{8}^{SC}&=&P_{\Pi},
\end{eqnarray}
where the superscript $SC$ stands for second-class. 
At this stage in the analysis, since we have the solutions for the Lagrange multipliers given previously by \eqref{lambda_1.1}, \eqref{lambda_2.1}, \eqref{lambda_3.1}, \eqref{lambda_5.1}, \eqref{lambda_6.1}, and \eqref{sol_lambda_7}, the total Hamiltonian can be derived by substituting these solutions into the original Hamiltonian \eqref{primary_hamiltonian}, leading to

\begin{eqnarray}
    H_{T}&=&H'+\lambda_4 C_{4}^{FC}+\lambda_{8}C_2^{FC},
\end{eqnarray}
with
\begin{widetext}
    \begin{eqnarray}
     H'&=&-\frac{N\Lambda \Pi^{1/2}}{\beta^2 +6\Pi^2 N^2 \Lambda}C_1^{FC}+(\frac{2}{\beta}m_0 kN^3 P_A+\frac{3}{2}n)C_3^{FC} \nonumber\\
    &+&\frac{N\Lambda \Pi^{1/2}}{\beta^2 +6\Pi^2 N^2\Lambda}\bigg[ Q(2\Pi^{1/2}C_8^{SC}+C_5^{SC}\beta\Pi^{-1/2})-\frac{\beta}{N}C_7^{SC}+C_6^{SC} \bigg]+C_8^{SC}p+C_7^{SC}\beta \nonumber\\
    &+& (4m_0 kN^2 P_A-6m_0 k\Pi^{1/2}NC_{3}^{SC}+\frac{1}{2}\frac{p\beta}{\Pi})C_{5}^{SC}+\frac{3}{4}\frac{\Pi^{1/2}}{N}(C_3^{SC})^2+\frac{n\beta}{4m_0 kN^2}C_3^{SC} \nonumber\\
    &+& (P_Q-6\Pi)C_4^{SC}-(2N\Pi^{1/2}Q+p)(C_8^{SC}+\frac{1}{8m_0 k}\frac{\beta}{N^2\Pi}C_{3}^{SC}) \nonumber\\
    &+&\bigg[ \frac{\beta}{2}\left( \frac{n}{N}-\frac{p}{\Pi} \right)-2m_0 kN^2 P_A \bigg](C_5^{SC}-\frac{1}{4m_0 k}C_3^{SC}) \nonumber\\
    &+&\bigg( \frac{2}{\beta}m_0 k N^3 P_A+\frac{1}{2}n \bigg)\bigg( \frac{1}{4m_0 k}\frac{\beta}{N^3}C_3^{SC}-\frac{\beta}{N}C_5^{SC} \bigg),
\end{eqnarray}
\end{widetext}
as the first-class Hamiltonian. We highlight that this constitutes a gauge theory, since $H_T$ is formed by constraints.
\subsection{Case $\alpha\approx0$ and $\beta\approx 0$.}
We have two secondary constraints denoted by $\psi_1:=\alpha$ and $\psi_2:=\beta$, which consistency relations give rise to the following expressions: 
\begin{eqnarray}
    \Dot{\psi}_1&=& \lambda_5, \\
    \Dot{\psi}_2&=& 4m_0 k N^2 P_A +\lambda_6.
\end{eqnarray}
Using the secondary constraints $\psi_1$ and $\psi_2$, the consistency relations collapse to the following:
\begin{eqnarray}
    \label{phi_1_a_b}\Dot{\phi}_1&=& -6(2N\Pi^{1/2}Q+p)-6\lambda_3, \\
    \Dot{\phi}_2&=&0, \\
    \label{phi_3_a_b}\Dot{\phi}_3&=& -3N\Pi^{-1/2}Q^2 +3N\Pi^{1/2}\Lambda+6\lambda_1, \\
    \Dot{\phi}_4&=&0, \\
    \label{phi_5_a_b}\Dot{\phi}_5&=&\frac{1}{4m_0 kN^2}\lambda_6, \\
    \label{phi_6_a_b}\Dot{\phi}_6&=&-P_B-6m_0 k A, \\
    \label{phi_7_a_b}\Dot{\phi}_7&=& -2\Pi^{1/2}(3Q^2-\Pi\Lambda), \\
    \Dot{\phi}_8&=&0, \\
    \label{psi_1_a_b}\Dot{\psi}_1&=& \lambda_5, \\
    \label{psi_2_a_b}\Dot{\psi}_2&=& 4m_0 kN^2 P_A+\lambda_6.
\end{eqnarray}
From \eqref{phi_1_a_b}, \eqref{phi_3_a_b}, \eqref{phi_5_a_b} and \eqref{psi_1_a_b} we find the following relations for the Lagrange multipliers correspondingly,
\begin{eqnarray}
    \label{lambda_3_2}\lambda_3&=&-(2N\Pi^{1/2}Q+p), \\
    \label{lambda_1_2}\lambda_1&=& \frac{1}{2}N\Pi^{-1/2}Q^2-\frac{1}{2}N\Pi^{1/2}\Lambda, \\
    \label{lambda_6_2}\lambda_6&=&0, \\
    \label{lambda_5_2}\lambda_5&=& 0.
\end{eqnarray}
 On the other hand, by the relations \eqref{phi_6_a_b}, \eqref{phi_7_a_b} and \eqref{psi_2_a_b}, and using \eqref{lambda_6_2}, we notice that there are no restrictions on any Lagrange multiplier. Thus, the following tertiary constraints arise:
\begin{eqnarray}
    \chi_1&=& P_B+6m_0 k A, \\
  \label{chi_2_second_case}  \chi_2&=& 3Q^2 -\Pi\Lambda, \\
    \chi_3&=& P_A.
\end{eqnarray}
There is a correlation between the Lagrange multiplier $\lambda_1$ and the constraint $\chi_2$ given by:

\begin{equation}
    \chi_2=6\frac{\Pi^{1/2}}{N}\lambda_1+2\Pi\Lambda.
\end{equation}
The consistency relations
corresponding to the tertiary constraints result in the following:
\begin{eqnarray}
    \Dot{\chi}_1&=& 24m_0^2 k^2 N^2 \phi_6+6m_0 kN \psi_1 \approx0, \\
    \Dot{\chi}_2&=& 6Qq, \\
    \Dot{\chi}_3&=& -6m_0 k \psi_2 \approx 0.
\end{eqnarray}
We observe that $\Dot{\chi}_2$ leads to quaternary constraints with three distinct scenarios to consider:\\

i) A quaternary constraint emerges when $Q\approx 0$ with $q\neq 0$.\\

ii) A quaternary constraint occurs when $q\approx 0$, allowing $Q\neq 0$ as a free variable.\\

iii) Give rise to two quaternary constraints $Q\approx0$ and $q\approx0$.\\

The subsequent discussion will focus on cases ii) and iii), as these are the only cases that demonstrate consistency. Case i) is determined to be inconsistent, as detailed in Appendix~\ref{non_permissible_Palatini}.
\subsubsection{Case $q\approx0$ and $Q\neq 0$.}\label{Case_3_permissible}
The quaternary constraint is denoted by $\xi:=q$, whose corresponding consistency relation reads
\begin{equation}
    \Dot{\xi}= \lambda_2.
\end{equation}
This result gives rise to the following condition for the Lagrange multiplier $\lambda_2$ given by
\begin{equation}\label{lambda_2_2}
    \lambda_2=0.
\end{equation}
The procedure ends here.

Furthermore, for this case, it is worth highlighting that $\Lambda\neq0$. This result comes from the expression \eqref{chi_2_second_case} and from the fact that $Q\neq0$. 

Bearing in mind the secondary constraints $\psi_1=\alpha$ and $\psi_2=\beta$, the complete set of constraints can be rewritten as follows:
\begin{eqnarray}
    \label{Cons_1_alpha_beta}\phi_1&=& P_Q-6\Pi,\\
    \phi_2&=& P_q, \\
    \phi_3&=& P_{\Pi}, \\
    \phi_4&=& P_{p}, \\
    \phi_5&=& P_{\alpha}, \\
    \phi_6&=& P_{\beta}, \\
    \phi_7&=& P_N, \\
    \phi_8&=& P_n,\\
    \psi_1&=& \alpha, \\
    \psi_2&=& \beta, \\
    \chi_1&=& P_B+6m_0 k A, \\
    \chi_2&=& 3Q^2 -\Pi\Lambda, \\
    \chi_3&=& P_A, \\
    \label{Cons_14_alpha_beta}\xi&=& q.
\end{eqnarray}
Computing the Poisson bracket among the constraints \eqref{Cons_1_alpha_beta}-\eqref{Cons_14_alpha_beta} to find a complete set of constraints of first-class and second-class constraints, we get:

\begin{widetext}
\begin{equation}\label{matrix_form_2}
[C_A,C_B]=
\begin{pmatrix}
0 & 0 & -6&0 &0&0&0&0&0&0&0&-6Q&0&0\\
0 & 0 & 0&0&0&0&0&0&0&0&0&0&0&-1\\
6 & 0 & 0&0&0&0&0&0&0&0&0&\Lambda&0&0 \\
0&0&0& 0&0&0&0&0&0&0&0&0&0&0 \\
0&0&0&0&0&0&0&0&-1&0&0&0&0&0 \\
0&0&0&0&0&0&0&0&0&-1&0&0&0&0 \\
0&0&0&0&0&0&0&0&0&0&0&0&0&0 \\
0&0&0&0&0&0&0&0&0&0&0&0&0&0 \\
0&0&0&0&1&0&0&0&0&0&0&0&0&0 \\
0&0&0&0&0&1&0&0&0&0&0&0&0&0 \\
0&0&0&0&0&0&0&0&0&0&0&0&6m_0 k&0 \\
6Q&0&-\Lambda&0&0&0&0&0&0&0&0&0&0&0\\
0&0&0&0&0&0&0&0&0&0&-6m_0 k&0&0&0\\
0&1&0&0&0&0&0&0&0&0&0&0&0&0\\
\end{pmatrix} \,,
\end{equation}
\end{widetext}
where the indices $A$ and $B$ run from $1$ to $14$. Each number indicates the constraints $\phi_1$, $\phi_2$, $\phi_3$, $\phi_4$, $\phi_5$, $\phi_6$, $\phi_7$, $\phi_8$, $\psi_1$, $\psi_2$, $\chi_1$, $\chi_2$, $\chi_3$ and $\xi$ respectively. The rank of the matrix is $10$, with nullity $14-10=4$. Hence, there are four independent eigenvectors whose eigenvalue vanishes. This suggests that there are four independent combinations of constraints that are first-class:
\begin{eqnarray}
    \label{C_1_FC_2}C_1^{FC}&=& -\frac{\Lambda}{6}P_Q+QP_{\Pi}+3Q^2, \\
    \label{C_2_FC_2}C_2^{FC}&=&P_p, \\
    \label{C_3_FC_2}C_3^{FC}&=& P_N, \\
    \label{C_4_FC_2}C_4^{FC}&=&P_n.
\end{eqnarray}
To continue our analysis, we express the $10$ second-class constraints as follows:
\begin{eqnarray}
    \label{C_1_SC_q_vann}C_1^{SC}&=&P_q, \\
    C_{2}^{SC}&=&P_{\alpha}, \\
    C_{3}^{SC}&=& P_{\beta}, \\
    C_{4}^{SC}&=& \alpha, \\
    C_{5}^{SC}&=& \beta, \\
    C_{6}^{SC}&=&P_{B}+6m_0 k A, \\
    C_{7}^{SC}&=& P_A, \\
    C_{8}^{SC}&=& q, \\
    C_{9}^{SC}&=& P_{Q}-6\Pi, \\
    \label{C_10_SC_q_vann}C_{10}^{SC}&=& 3Q^2-\Pi \Lambda.
\end{eqnarray}
At this stage, knowing that the solutions for the Lagrange multipliers for this case are given by \eqref{lambda_3_2}-\eqref{lambda_5_2} and \eqref{lambda_2_2}, the total Hamiltonian $H_T$ becomes
\begin{equation}
    H_{T}=H'+\lambda_4 C_2^{FC}+\lambda_7C_3^{FC}+\lambda_8C_4^{FC},
\end{equation}
with
\begin{widetext}
    \begin{eqnarray}
    H'&=& N\Pi^{1/2}C_1^{FC}+nC_3^{FC}+C_9^{SC}C_8^{SC}+N\Pi^{1/2}\left( \frac{\Lambda}{6}C_9^{SC}+C_{10}^{SC} \right)-\frac{(2N\Pi^{1/2}Q+p)}{8m_0 kN^2 \Pi}C_4^{SC}C_5^{SC} \nonumber\\
    &+&C_5^{SC}C_6^{SC}+4m_0 kN^2 C_3^{SC}C_7^{SC}-6m_0k\Pi^{1/2}NC_4^{SC}C_3^{SC}+\frac{1}{2}\frac{p}{\Pi}C_5^{SC}C_3^{SC}+\frac{n}{N}C_5^{SC}C_3^{SC} \nonumber\\
    &+&\frac{3}{4}\frac{\Pi^{1/2}}{N}(C_4^{SC})^2+\frac{\Pi^{-1/2}}{N}(C_5^{SC})^2+\frac{n}{4m_0kN^3}C_4^{SC}C_5^{SC}+\frac{1}{2}N\Pi^{-1/2}(Q^2-\Pi\Lambda)C_9^{SC},
\end{eqnarray}
\end{widetext}
as the first-class Hamiltonian. As expected, this is a gauge theory since $H_T$ is a combination of constraints.

\subsubsection{Case $Q\approx0$ and $q\approx0$.}\label{Case_4_permissible}
We have two quaternary constraints, denoted by $\xi_1:=Q$ and $\xi_2:=q$. The consistency relations are expressed as:
%\end{enumerate}
\begin{eqnarray}
    \Dot{\xi}_1&=&-\frac{1}{2}\Pi^{1/2}\Lambda, \\
    \Dot{\xi}_2&=& \lambda_2.
\end{eqnarray}
The first relation implies that $\Lambda=0$ since $\Pi\neq0$, and the second gives us the following condition for the Lagrange multiplier $\lambda_2$ given by,
\begin{equation}\label{lambda_2_3}
    \lambda_2=0.
\end{equation}
This result, together with the fact that $Q\approx0$, the solution of the Lagrange multipliers given above by \eqref{lambda_1_2} and \eqref{lambda_3_2} collapses to
\begin{eqnarray}
\label{lambda_1_3}\lambda_1&=&0, \\
    \label{lambda_3_3}\lambda_3&=&-p.
\end{eqnarray}
The process of identifying constraints concludes at this stage, since all constraints involved are preserved weakly in time, and there are no additional conditions on the Lagrange multipliers. It is important to emphasize that this scenario is only valid when $\Lambda=0$; otherwise, it becomes inconsistent. Thus, bearing in mind the secondary constraints $\psi_1=\alpha$ and $\psi_2=\beta$, the complete set of constraints can be rewritten as follows:

\begin{eqnarray}
    \label{con_1_alpha_beta_q_Q_van}\phi_1&=& P_Q-6\Pi,\\
    \phi_2&=& P_q, \\
    \phi_3&=& P_{\Pi}, \\
    \phi_4&=& P_{p}, \\
    \phi_5&=& P_{\alpha}, \\
    \phi_6&=& P_{\beta}, \\
    \phi_7&=& P_N, \\
    \phi_8&=& P_n, \\
    \psi_1&=& \alpha,\\
    \psi_2&=& \beta,\\
    \chi_1&=& P_{B}+6m_0 k A, \\
    \chi_2&=& Q, \\
    \chi_3&=& P_A, \\
    \label{con_14_alpha_beta_q_Q_van}\xi&=&q.
\end{eqnarray}
%\end{widetext}
To continue our analysis, we compute the Poisson brackets between each constraint given in \eqref{con_1_alpha_beta_q_Q_van}-\eqref{con_14_alpha_beta_q_Q_van} to determine the first-class and second-class constraints present in this scenario, giving as a result:

\begin{widetext}
    \begin{equation}\label{matrix_form_3}
[C_A,C_B]=
\begin{pmatrix}
0 & 0 & -6&0 &0&0&0&0&0&0&0&-1&0&0\\
0 & 0 & 0&0&0&0&0&0&0&0&0&0&0&-1\\
6 & 0 & 0&0&0&0&0&0&0&0&0&0&0&0 \\
0&0&0& 0&0&0&0&0&0&0&0&0&0&0 \\
0&0&0&0&0&0&0&0&-1&0&0&0&0&0 \\
0&0&0&0&0&0&0&0&0&-1&0&0&0&0 \\
0&0&0&0&0&0&0&0&0&0&0&0&0&0 \\
0&0&0&0&0&0&0&0&0&0&0&0&0&0 \\
0&0&0&0&1&0&0&0&0&0&0&0&0&0 \\
0&0&0&0&0&1&0&0&0&0&0&0&0&0\\
0&0&0&0&0&0&0&0&0&0&0&0&6m_0 k&0\\
1&0&0&0&0&0&0&0&0&0&0&0&0&0\\
0&0&0&0&0&0&0&0&0&0&-6m_0 k&0&0&0\\
0&1&0&0&0&0&0&0&0&0&0&0&0&0
\end{pmatrix} \,,
\end{equation}
\end{widetext}
where the indices $A$ and $B$ run from $1$ to $14$. Each number represents the constraint $\phi_1$, $\phi_2$, $\phi_3$, $\phi_5$, $\phi_6$, $\phi_7$, $\phi_8$, $\psi_1$, $\psi_2$, $\chi_1$, $\chi_2$, $\chi_3$, and $\xi$ respectively. The rank of this $14\times14$ matrix is $10$, with nullity $14-10=4$. Therefore, there are four independent eigenvectors whose corresponding eigenvalue vanishes. This result suggests that there are $4$ independent combinations of constraints that are first-class given by
\begin{eqnarray}
    C_1^{FC}&=& -\frac{1}{6}P_{\Pi}+Q, \\
    C_2^{FC}&=& P_p, \\
    C_3^{FC}&=&P_N, \\
    C_4^{FC}&=& P_n.
\end{eqnarray}
Thus, the remaining $10$ independent second-class constraints become

\begin{eqnarray}   \label{SC1_Q_q_vanish}C_1^{SC}&=&P_{Q}-6\Pi, \\
\label{SC2_Q_q_vanish}C_{2}^{SC}&=& P_q, \\
\label{SC3_Q_q_vanish}C_{3}^{SC}&=& P_{\alpha}, \\
    \label{SC4_Q_q_vanish}C_4^{SC}&=&P_{\beta}, \\
    \label{SC5_Q_q_vanish}C_5^{SC}&=&\alpha, \\
  \label{SC6_Q_q_vanish}C_6^{SC}&=&\beta, \\
    \label{SC7_Q_q_vanish}C_7^{SC}&=& P_{B}+6m_0 kA, \\
    \label{SC8_Q_q_vanish}C_8^{SC}&=&Q, \\
    \label{SC9_Q_q_vanish}C_9^{SC}&=&P_A, \\
    \label{SC_10_Q_q_vanish}C_{10}^{SC}&=& q.
\end{eqnarray}
Taking into account the solutions of the Lagrange multipliers given by \eqref{lambda_2_3}-\eqref{lambda_3_3}, \eqref{lambda_6_2} and \eqref{lambda_5_2}, the total Hamiltonian reads
\begin{equation}
H_T=H'+\lambda_4C_2^{FC}+\lambda_7C_3^{FC}+\lambda_8C_4^{FC},
\end{equation}
with
\begin{widetext}
\begin{eqnarray}
H'&=&C_1^{FC}C_{10}^{SC}+nC_3^{FC}+C_6^{SC}C_7^{SC}+4m_0k N^2P_AC_4^{SC}-6m_0k\Pi^{1/2}NC_4^{SC}C_5^{SC}+\frac{1}{2}\frac{p}{\Pi}C_4^{SC}C_6^{SC}\nonumber\\
    &+&\frac{n}{N}C_4^{SC}C_6^{SC}+\frac{3}{4}\frac{\Pi^{1/2}}{N}(C_5^{SC})^{2}+\frac{\Pi^{-1/2}}{N}(C_6^{SC})^2+\frac{1}{4m_0k}\frac{n}{N^3}C_5^{SC}C_6^{SC}+6N\Pi^{1/2}(C_8^{SC})^2 \nonumber\\
    &-&\frac{p C_5^{SC}C_6^{SC}}{8m_0 kN^2 \Pi},
\end{eqnarray}
\end{widetext}
as the first-class Hamiltonian. At this stage, we can recognize that it is a gauge theory, as the total Hamiltonian consists of constraints.\\

\section{First-class Hamiltonian}\label{sec_A_1}
The theories outlined in \ref{Case_1_permissible}, \ref{Case_2_permissible}, \ref{Case_3_permissible}, and \ref{Case_4_permissible} are gauge theories because each of their respective extended Hamiltonians is formed solely by constraints. 
All of them involve second-class constraints, which must be either handled with the Dirac bracket~\cite{dirac1964lectures} or explicitly solved. We will follow the second way. \\

\subsection{Solution to the second-class constraints}\label{subsection_A}
In what follows, we solve explicitly the sets of second-class constraints in terms of pairs of canonical variables of each permissible case.

\subsubsection{Case $\alpha\approx0$, $\beta\neq0$, with $Q\approx0$ and $q\neq 0$}\label{Consistent_Sol1}
From the set of second-class constraints given by \eqref{C1SC_Q_van}-\eqref{C10SC_Q_van}, there are $4$ nontrivial constraints to solve given by \eqref{C1SC_Q_van}, \eqref{C7SC_Q_van}, \eqref{C8SC_Q_van}, and \eqref{C10SC_Q_van}, which can be solved straightforwardly in terms of the following $4$ canonical coordinates $P_Q$, $q$, $\Pi$, and $N$, leading to
\begin{eqnarray}
    P_Q&=& -\frac{3}{4\Lambda}(P_B+6m_0kA)^2, \\
    q&=&\frac{2}{3}\frac{\beta}{(P_B+6m_0kA)}, \\
    \Pi&=&-\frac{1}{8\Lambda}(P_B+6m_0kA)^2, \\
    \label{solution_N}N&=&\frac{4\beta}{(P_B+6m_0kA)^2}\sqrt{-2\Lambda}.
\end{eqnarray}
Hence, only first-class constraints remain, and the extended action reads
\begin{eqnarray}\label{ext_action_1}
    S_E&=& \int \bigg[P_p\Dot{p}+P_A\Dot{A}+P_B\Dot{B}+P_{\beta}\Dot{\beta}+P_n\Dot{n} \nonumber\\
    &-&\mu_1C_1^{FC}-\mu_2C_2^{FC}-\mu_3C_3^{FC}\bigg]dt,
\end{eqnarray}
with
\begin{eqnarray}
    C_1^{FC}&=& P_p, \\
    C_2^{FC}&=& P_n, \\
    \label{C_3_F_C_first_formulation}C_3^{FC}&=& \frac{P_\beta(P_B+6m_0kA)^2}{4\sqrt{-2\Lambda}},
\end{eqnarray}
and $\mu_1$, $\mu_2$, and $\mu_3$ as Lagrange multipliers. 
Therefore, the phase space is labeled by the following $5$ pairs of canonical coordinates $(p,P_p)$, $(A,P_A)$, $(B,P_B)$, $(\beta,P_{\beta})$ and $(n,P_n)$ with $3$ first-class constraints.
From \eqref{C_3_F_C_first_formulation}, we deduce that $\Lambda <0$ is necessary for a real formulation. However, it is worth stressing that, through a Darboux transformation~\cite{Arnold_book}, it is possible to find other sets of canonical coordinates to describe the phase space to give rise to a real formulation with a cosmological constant $\Lambda>0$.

We confirm that this formulation has $2$ local degrees of freedom; namely:
\begin{equation}
    \mathrm{D.O.F}=\frac{1}{2}[10-2\times3]=2.
\end{equation}

\subsubsection{Case $\alpha\approx0$, $\beta\neq0$, with $Q\neq0$ and $q\approx0$}\label{Consistent_Sol2}

The set of second-class constraints of this scenario is provided by \eqref{SC1.1}-\eqref{SC8.1}, wherein $C_6^{SC}$ and $C_7^{SC}$ are solved for $\Pi$ and $\beta$ as unknowns to obtain canonical coordinates. At the same time, the rest are trivial to solve since they correspond to coordinates of the phase space, namely:

\begin{eqnarray}    \label{Pi_solution_SC}\Pi&=&\frac{3Q^2}{\Lambda}-\frac{(\frac{1}{2}P_B+3m_0 k A)^2}{2\Lambda}, \\  \label{beta_solution_SC}\beta&=& -\frac{N}{\sqrt{|\Lambda|}}(\frac{1}{2}P_B+3m_0 kA) \times \nonumber\\
    &&\bigg[ 3Q^2 -\frac{(\frac{1}{2}P_B+3m_0 k A)}{2} \bigg]^{1/2},
\end{eqnarray}
with $\Lambda\neq 0$. Nevertheless, analogously to the first case, through a Darboux transformation~\cite{Arnold_book}, it is possible to obtain other sets of canonical coordinates to give rise to a formulation with or without cosmological constant $\Lambda$. Thus, in terms of the canonical variables $(Q,P_Q)$, $(p,P_p)$, $(A,P_A)$, $(B,P_B)$, $(N,P_N)$ and $(n,P_n)$, the extended action related to \eqref{action1} becomes:
\begin{widetext}
\begin{eqnarray}\label{ext_action_2}
    S_E&=&\int \bigg[ P_Q \Dot{Q} +P_p \Dot{p}+P_A \Dot{A}+P_B \Dot{B}+P_N \Dot{N}+P_n \Dot{n} -\mu_1C_1^{FC}-\mu_2 C_2^{FC}-\mu_3C_{3}^{SC}-\mu_4C_4^{FC} \bigg]dt,\nonumber\\
\end{eqnarray}
with
\begin{eqnarray}
    C_1^{FC}&=& \frac{N^2}{\Lambda}\bigg[ P_Q-\frac{6}{\Lambda}\bigg( 3Q^2-\frac{(\frac{1}{2}P_B+3m_0 k A)^2}{2}
 \bigg) \bigg]\bigg[ 3Q^2-\bigg( \frac{1}{2}P_B+3m_0 k A \bigg)^2 \bigg]\times \nonumber\\
 &&\bigg[ \bigg( \frac{1}{2}P_B+3m_0 k A \bigg)^2+6\bigg( 3Q^2-\frac{(\frac{1}{2}P_B+3m_0 k A)^2}{2} \bigg) \bigg] \quad\quad\quad\quad \mathrm{with}\quad\Lambda\neq 0, \\
 C_2^{FC}&=&P_n, \\
 C_3^{FC}&=& P_N, \\
 C_{4}^{FC}&=& P_p.
\end{eqnarray}
 \end{widetext}
This formulation is given by $12$ canonical coordinates $(Q,P_Q)$, $(p,P_p)$, $(A,P_A)$, $(B,P_B)$, $(N,P_N)$ and $(n,P_n)$ and $4$ first-class constraints $C_1^{FC}$, $C_2^{FC}$, $C_3^{FC}$ and $C_4^{FC}$, implying that the number of degrees of freedom becomes:
\begin{equation}
    \mathrm{D.O.F.}=\frac{1}{2}[12-2\times4]=2.
\end{equation}

\subsubsection{Case $\alpha\approx0$, $\beta\approx0$, with $Q\neq0$ and $q\approx0$}\label{Consistent_Sol3}
The set of second-class constraints is given by \eqref{C_1_SC_q_vann}-\eqref{C_10_SC_q_vann}, where the nontrivial to solve are $C_6^{SC}$, $C_9^{SC}$ and $C_{10}^{SC}$. A possible solution to obtain canonical coordinates is given by the following:  
\begin{eqnarray}
\label{A_2}A&=&-\frac{P_B}{6m_0k},\\
\label{Solution_PQ_2}P_Q&=&6\Pi, \\
    \label{Solution_Q_2}Q&=& \left( \bigg|\frac{\Pi \Lambda}{3}\bigg| \right)^{1/2}, \quad\quad \mathrm{with}\quad\Lambda>0.
\end{eqnarray}
Plugging \eqref{A_2}, \eqref{Solution_PQ_2}, and \eqref{Solution_Q_2} into the extended action related to \eqref{action1} yields the following:

    \begin{eqnarray}\label{ext_action_3}
    S_E=\int\bigg[ P_{\Pi}\Dot{\Pi}+P_p\Dot{p}+P_B \Dot{B}+P_N \Dot{N}+P_n\Dot{n} \nonumber \\-\mu_1 C_1^{FC}-\mu_2C_2^{FC}-\mu_3C_3^{FC}-\mu_4C_4^{FC} \bigg] dt, \nonumber \\
\end{eqnarray}
where the first-class constraints \eqref{C_1_FC_2}-\eqref{C_4_FC_2} are rewritten in terms of the canonical pair $(\Pi,P_{\Pi})$, $(p,P_p)$, $(B,P_B)$, $(N,P_N)$ and $(n,P_n)$ as follows:
\begin{eqnarray}
    C_1^{FC}&=&-P_{\Pi}\left(\bigg| \frac{\Pi\Lambda}{3}\bigg| \right)^{1/2} \; \; \mathrm{with}\; \; \Lambda>0, \\
    C_2^{FC}&=& P_p, \\
    C_3^{FC}&=& P_N, \\
    C_4^{FC}&=& P_n.
\end{eqnarray}
It is worthwhile to note that the presence of the cosmological constant $\Lambda$ can vanish, as we can redefine the following Lagrange multiplier $\nu_1:=-\mu_1\left( \frac{\Pi\Lambda}{3} \right)^{1/2}$. This implies that the associated first-class constraints reduce to
\begin{equation}
    C_1^{'FC}:=P_{\Pi}.
\end{equation}
Therefore, the cosmological constant $\Lambda$ does not appear in this canonical formulation. Thus, the number of degrees of freedom is given by
\begin{eqnarray}
    \mathrm{D.O.F}&=&\frac{1}{2}[10-2\times 4] \nonumber\\
    &=& 1.
\end{eqnarray}

\subsubsection{Case $\alpha\approx0$, $\beta\approx0$, with $Q\approx0$ and $q\approx0$}\label{Consistent_Sol4}
In this case, the set of second-class constraints is given by \eqref{SC1_Q_q_vanish}-\eqref{SC_10_Q_q_vanish}, where the non-trivial ones to solve are given by \eqref{SC1_Q_q_vanish} and \eqref{SC7_Q_q_vanish}, which correspond to $C_1^{SC}$ and $C_7^{SC}$, respectively. A possible solution to such constraints is given by
\begin{eqnarray}
    P_Q&=&6\Pi, \\
    A&=&-\frac{P_B}{6m_0k},
\end{eqnarray}
where the phase space is labeled by the following canonical coordinates $(\Pi, P_{\Pi})$, $(p,P_p)$, $(B,P_B)$, $(N,P_N)$, and $(n,P_n)$. Hence, the extended action corresponding to \eqref{action1} becomes
\begin{eqnarray}\label{ext_action_4}
S_E=\int_M[P_{\Pi}\Dot{\Pi}+P_p\Dot{p}+P_B\Dot{B}+P_N\Dot{N}+P_n\Dot{n} \nonumber \\ -\mu_1C_1^{FC}-\mu_2C_2^{FC}-\mu_3C_3^{FC}-\mu_4C_4^{FC}]dt, 
    \end{eqnarray}

with
\begin{eqnarray}
    C_1^{FC}&=&-\frac{1}{6}P_{\Pi}, \\
    C_2^{FC}&=&P_p, \\
    C_3^{FC}&=&P_N, \\
    C_4^{FC}&=& P_n.
\end{eqnarray}
Redefining the following Lagrange multiplier as $\nu_1:=-\mu_1/6$, we can realize that the canonical formulation \eqref{ext_action_4} is the same as given above in \eqref{ext_action_3} since the canonical coordinates that describe the phase space along with the set of first-class constraints coincide. Therefore, since \eqref{ext_action_3} and \eqref{ext_action_4} are valid for $\Lambda>0$ and $\Lambda=0$ respectively, we can confirm that the formulations they represent can be derived regardless of the value of the cosmological constant as long as $\Lambda \geq 0$. Thus, the number of degrees of freedom is:
\begin{eqnarray}
    \mathrm{D.O.F}&=&\frac{1}{2}[10-2\times 4] =1.
\end{eqnarray}
We close this section with the following five remarks:\\

(i) Because the cases \ref{Consistent_Sol3} and \ref{Consistent_Sol4} collapse to the same formalism when the second-class constraints are solved, we can confirm that the Hamiltonian analysis of \eqref{action1} acquires $3$ distinct and consistent scenarios given by \eqref{ext_action_1}, \eqref{ext_action_2}, and \eqref{ext_action_3} [or \eqref{ext_action_4}]. \\ 

(ii) When solving second-class constraints in scenarios \ref{Consistent_Sol1} and \ref{Consistent_Sol2}, the cosmological constant $\Lambda$ does not vanish. Specifically, $\Lambda<0$ for case \ref{Consistent_Sol1}  and $\Lambda>0$ for case \ref{Consistent_Sol2}, ensuring a real formulation. In contrast to scenarios \ref{Consistent_Sol3} and \ref{Consistent_Sol4}, the resolution of second-class constraints does not involve the presence of the cosmological constant $\Lambda$.\\

(iii) The Hamiltonian analysis of the coupled action \eqref{action1} can not have a uniform treatment, since the system leads to three different consistent scenarios given by \eqref{ext_action_1}, \eqref{ext_action_2}, and \eqref{ext_action_3} [or \eqref{ext_action_4}]. This stems from the fact that the consistency relation of the constraint \eqref{8_prim_constraint} gives rise to several cases that need to be considered. Hence, the theory \eqref{action1} is somewhat pathological~\cite{HenneauxTeitelboim+1992}. \\

(iv) According to Appendix~\ref{canonical_analysis_Palatini}, which discusses the analysis of the $4$-dimensional Palatini action considering homogeneity and isotropy in the fields involved, the pair $(Q,\Pi)$ describes the phase space. Here, $\Pi$ denotes the canonical momentum associated with the canonical coordinate $Q$. Unlike the three different scenarios outlined in this work, detailed in formulations \eqref{ext_action_1}, \eqref{ext_action_2}, and \eqref{ext_action_3} [or \eqref{ext_action_4}], this role is not being fulfilled. Moreover, the situations differ as follows: in the first scenario, there are no $Q$ and $\Pi$; in the second scenario, $Q$ serves as a coordinate with its canonical momentum denoted as $P_Q$, while no $\Pi$ is involved; and in the third scenario, there is no $Q$, and $\Pi$ acts as a coordinate with its canonical momentum $P_{\Pi}$. \\

(v) After solving the second-class constraints, the extended action \eqref{ext_action_1} does not involve the presence of the lapse function $N$, while in the scenarios \eqref{ext_action_2}, and \eqref{ext_action_3} [or \eqref{ext_action_4}], the lapse function $N$ serves as a canonical coordinate, with its corresponding canonical momentum defined as $P_N$. Unlike the canonical analysis of the $4$-dimensional Palatini action (see Appendix~\ref{canonical_analysis_Palatini}), the lapse function is a Lagrange multiplier because it does not involve any time derivatives.

%%%%%%%%%%%%%%%%%%%%%%%%%%%%%%%%%%%%%%%%%%

\section{Canonical analysis of 4-dimensional effective theory}\label{sec_V}
To enrich the content of this work, in this section, we perform the canonical analysis of the action \eqref{eff_action} by focusing solely on it. We consider the relationship between the metric and the orthonormal field \eqref{relation_metric_tethrad} together with the parameterization of the tetrad field \eqref{e_t_I} and \eqref{e_a_I} as well as the properties of homogeneity and isotropy in all the fields involved. The action simplifies to:
\begin{eqnarray}
    S_{Eff}&=&\int\bigg[-6m_0 k A \Dot{B}+\frac{3}{4}\frac{\Pi^{1/2}}{N}\Dot{A}^2-\frac{\Pi^{-1/2}}{N}\Dot{B}^2\nonumber\\
    &+&\frac{\Pi^{1/2}}{4m_0 k}\Dot{A}\bigg(-\frac{1}{2N}\Pi^{-3/2}\Dot{\Pi}\Dot{B}+\frac{1}{N}\Pi^{-1/2}\Ddot{B} \nonumber\\
    &-&\frac{\Dot{N}}{N^2}\Pi^{-1/2}\Dot{B}\bigg)\bigg] dt.
\end{eqnarray}
Since the action involves second-time derivatives, we follow the Ostrogradski methodology. To establish the Hamiltonian description of the above higher-order action, we define all the canonical momenta involved here as follows:

\begin{eqnarray}
    && P_{\Pi}=\frac{\delta L}{\delta \Dot{\Pi}}=\frac{\partial L}{\partial \Dot{\Pi}}=-\frac{1}{8m_0 k}\frac{\Dot{A}\Dot{B}}{N^2 \Pi} \,,\\
    && P_{p}=\frac{\delta L}{\delta \Ddot{\Pi}}=\frac{\partial L}{\partial \Ddot{\Pi}}=0 \,, \\
    && P_A= \frac{\delta L}{\delta \Dot{A}}=\frac{\partial L}{\partial \Dot{A}}=\frac{3}{2}\frac{\Pi^{1/2}}{N}\Dot{A}\nonumber \\    
    &&+\frac{\Pi^{1/2}}{4m_0 kN}\bigg( -\frac{1}{2N}\Pi^{-3/2}\Dot{\Pi}\Dot{B}+\frac{1}{N}\Pi^{-1/2}\Ddot{B}\nonumber \\ && -\frac{\Dot{N}}{N^2}\Pi^{-1/2}\Dot{B} \bigg),\\  
   && P_{\alpha}=\frac{\delta L}{\delta \Ddot{A}}=\frac{\partial L}{\partial \Ddot{A}}=0 \,, 
   \end{eqnarray}
   \begin{eqnarray}
    && P_B=\frac{\delta L}{\delta \Dot{B}}=\frac{\partial L}{\partial \Dot{B}}-\frac{d}{dt}\left( \frac{\partial L}{\partial \Ddot{B}} \right)=\nonumber \\&& -6m_0 A-\frac{2}{N}\Pi^{-1/2}\Dot{B}-\frac{\Dot{A}\Dot{\Pi}}{8m_0 kN^2}+\frac{1}{4m_0 k}\frac{\Dot{A}\Dot{N}}{N^3} \nonumber \\&& -\frac{1}{4m_0 k}\frac{\Ddot{A}}{N^2} \,, \\
    && P_{\beta}=\frac{\delta L}{\delta \Ddot{B}}=\frac{\partial L}{\partial \Ddot{B}}=\frac{1}{4m_0 k}\frac{\Dot{A}}{N^2} \,\\
    && P_{N}=\frac{\delta L}{\delta \Dot{N}}=\frac{\partial L}{\partial \Dot{N}}=-\frac{1}{4m_0 k}\frac{\Dot{A}\Dot{B}}{N^3} \,, \\
    && P_n=\frac{\delta L}{\delta \Ddot{N}}=\frac{\partial L}{\partial \Ddot{N}}=0 \,,
\end{eqnarray}
with $p:=\Dot{\Pi}$, $\alpha:=\Dot{A}$, $\beta:=\Dot{B}$, $n:=\Dot{N}$. This gives rise to the set of canonical variables with their respective canonical momentum given by $(\Pi, P_{\Pi})$, $(p,P_p)$, $(A, P_A)$, $(\alpha, P_{\alpha})$, $(B, P_B)$, $(\beta, P_{\beta})$, $(N, P_N)$ and $(n,P_n)$. 
The computation of the canonical momentum associated with each coordinate gives rise to the following set of primary constraints
\begin{eqnarray}
    \label{phi_1_effremov}&&\phi_1:=P_{\Pi}+\frac{1}{8m_0 k}\frac{\alpha \beta}{N^2 \Pi} \,, \\
    \label{phi_2_effremov}&&\phi_2:= P_p \,, \\
    \label{phi_3_effremov}&&\phi_3:=P_{\alpha} \,, \\
    \label{phi_4_effremmov}&&\phi_4:=P_{\beta}-\frac{1}{4m_0 k}\frac{\alpha}{N^2} \,, \\
    \label{phi_5_effremov}&&\phi_5:=P_N+\frac{1}{4m_0 k}\frac{\alpha \beta}{N^3} \,, \\
    \label{phi_6_effremov}&& \phi_6:= P_n \,.
\end{eqnarray}
Considering a Legendre transformation in conjunction with the set of primary constraints, the primary Hamiltonian assumes the subsequent form:
\begin{eqnarray}\label{primary_hamiltonian_Effective}
    H_{P}&=&P_{\Pi}p+P_{B}\beta +4m_0 kN^2 P_{\beta}P_{A}\nonumber\\
    &-&6m_0k\pi^{1/2}\alpha N P_{\beta}+\frac{1}{2}\frac{p\beta}{\Pi}P_{\beta}+\frac{n\beta}{N}P_{\beta} \nonumber\\
    &+& P_{N}n +6m_0 kA\beta+\frac{3}{4}\frac{\Pi^{1/2}}{N}\alpha^2 \nonumber\\
    &+&\frac{\Pi^{-1/2}}{N}\beta^2 +\frac{\alpha}{4m_0 k}\frac{n\beta}{N^3}+\lambda_1\phi_1+\lambda_2\phi_2\nonumber\\
    &+&\lambda_3\phi_3+\lambda_4\phi_4+\lambda_5\phi_5+\lambda_6\phi_6.
\end{eqnarray}
Applying the consistency relations to the set of primary constraints, we deduce the following result:
\begin{eqnarray}
    \label{Dot_phi_1_Effremov}\Dot{\phi}_1 &=& \left(3m_0 k \Pi^{-1/2}\alpha N+\frac{1}{2}\frac{p\beta}{\Pi^2}\right)P_{\beta} \nonumber\\
    &+&\frac{1}{2}\frac{\Pi^{-3/2}}{N}\beta^2 -\frac{9}{8}\frac{\alpha^2}{\Pi^{1/2}N}+ \frac{1}{2}\frac{\alpha P_A}{\Pi} \nonumber\\
    &-&\frac{1}{8m_0 k}\frac{\alpha \beta}{\Pi N^2}\left( \frac{n}{N}+\frac{p}{2\Pi} \right) \nonumber \\ &+& \frac{1}{8m_0 k}\frac{\beta}{N^2 \Pi}\lambda_3
    + \frac{1}{8m_0 k}\frac{\alpha}{N^2 \Pi}\lambda_4  \nonumber\\ &-& \frac{1}{4m_0 k}\frac{\alpha \beta}{\Pi N^3}\lambda_{5}, \\
    \Dot{\phi}_2&=& -\phi_1-\frac{\beta}{2\Pi}\phi_4 \approx0,\\
    \Dot{\phi}_{3}&=& -\frac{1}{4m_0 k}\frac{1}{N^2}\left[ \frac{\beta}{N}(n-\lambda_5)+\frac{\beta}{2}\lambda_1 -\lambda_4 \right],\nonumber \\
    \Dot{\phi}_4&=&-P_B-\frac{1}{2}\frac{pP_{\beta}}{\Pi}-\frac{n}{N}P_{\beta}-6m_0 kA\nonumber\\
    &-&\frac{2\Pi^{-1/2}\beta}{N}+\frac{1}{4m_0 k}\frac{\alpha n}{N^3} \nonumber \\ &-& \frac{1}{4m_0 k}\frac{\alpha}{N^2}\left( \frac{\lambda_1}{2\Pi}-\frac{\lambda_5}{N} \right) 
    -\frac{1}{4m_0 k}\frac{1}{N^2}\lambda_3, \nonumber \\
    \end{eqnarray}
    \begin{eqnarray}
    \label{dot_phi_5_Effremov}\Dot{\phi}_5 &=& \bigg( -8m_0 kNP_A+6m_0k\Pi^{1/2}\alpha+\frac{n\beta}{N^2} \bigg)P_{\beta}\nonumber\\
    &+&\frac{\Pi^{-1/2}}{N^2}\beta^{2}-\frac{3}{4}\frac{\Pi^{1/2}}{N^2}\alpha^2 \nonumber \\ &+& \frac{1}{4m_0 k}\frac{\alpha \beta}{N^3}\bigg( \frac{n}{N}+\frac{p}{2\Pi} \bigg) 
     \nonumber \\ &+& \frac{\alpha}{N}P_A+\frac{1}{4m_0 k}\frac{\alpha\beta}{N^3\Pi}\lambda_1\nonumber\\
    &+&\frac{1}{4m_0 k}\frac{\beta}{N^3}\lambda_3-\frac{1}{4m_0 k}\frac{\alpha}{N^3}\lambda_4, \\
       \label{dot_phi_6_Effective} \Dot{\phi}_6&=&-\frac{1}{4m_0 k}\frac{\alpha\beta}{N^3}-\left(\phi_5+\frac{\beta}{N}\phi_4\right).
\end{eqnarray}

In the latter expression, we notice the absence of any Lagrange multiplier condition. This indicates that a secondary constraint must be considered in the analysis. Given that non-degenerate gravity is regarded throughout the work, namely $N\neq 0$, the secondary constraint simplifies to $\alpha\beta\approx0$. This led us to investigate three possible scenarios:\\

i) A secondary constraint occurs when $\alpha\approx0$, allowing $\beta\neq0$ to be a free variable.\\

ii) A secondary constraint emerges when $\beta\approx 0$, while $\alpha\neq 0$ remains as a free variable.\\

iii) Two secondary constraints arise, represented by conditions $\alpha\approx0$ and $\beta\approx0$.\\

In contrast to action \eqref{action1}, only case (iii) proves to be consistent during the canonical analysis. Appendix~\ref{non_permissible_Effremov} discusses the cases (i) and (ii), which are not permissible to perform the canonical analysis. 

\subsection{Case $\alpha\approx0$ and $\beta\approx0$.}
We have two secondary constraints, which we denote by $\psi_1:=\alpha$ and $\psi_2:=\beta$. The consistency relations of the latter read as follows:
\begin{eqnarray}
    \label{Dot_psi_1_effremov}\Dot{\psi}_1&=&\lambda_3, \\
    \label{dot_psi_2_effremov}\Dot{\psi}_2&=& 4m_0 kN^2 P_A+\lambda_4.
\end{eqnarray}
Using the secondary constraints $\psi_1$ and $\psi_2$, the above set of consistency relations \eqref{Dot_phi_1_Effremov}-\eqref{dot_phi_6_Effective} collapses to
\begin{eqnarray}
    \Dot{\phi}_1&=& 0, \\
    \Dot{\phi}_2&=&0, \\
    \label{Dot_phi_3_Effremov_After}\Dot{\phi}_3&=& \frac{1}{4m_0 k N^2}\lambda_4, \\
    \label{Dot_phi_4_effrmov_after}\Dot{\phi}_4&=& -P_B-6m_0 kA -\frac{1}{4m_0 k}\frac{1}{N^3}\lambda_3,\\
    \Dot{\phi}_5&=& 0, \\
    \Dot{\phi}_6&=& 0. 
\end{eqnarray}
From \eqref{Dot_psi_1_effremov} and \eqref{Dot_phi_3_Effremov_After} we find that
\begin{eqnarray}
    \label{solution_lambda_3_Effremov}\lambda_3&=&0, \\
    \label{solution_lambda_4_Effremov}\lambda_4&=&0.
\end{eqnarray}
Given that $\lambda_3$ and $\lambda_4$ have already been solved, the consistency relations \eqref{Dot_phi_4_effrmov_after} and \eqref{dot_psi_2_effremov} yield two tertiary constraints expressed as: 
\begin{eqnarray}
    \chi_1&=& P_B+6m_0 k A, \\
    \chi_2&=& P_A.
\end{eqnarray}
The consistency relations to $\chi_1$ and $\chi_2$ result in the following:
\begin{eqnarray}
    \Dot{\chi}_1&=& 24 m_0^2 k^2 (\phi_4+\frac{1}{4m_0k N^2}\psi_1) \nonumber\\
    &\approx&0, \\
    \Dot{\chi}_2&=& 6m_0 k \psi_2\nonumber\\
    &\approx&0. 
\end{eqnarray}
The procedure of finding constraints ends at this stage. Thus, taking into account the secondary constraints $\psi_1=\alpha$ and $\psi_2=\beta$, the complete set of constraints can be rewritten as follows:

\begin{eqnarray}
    \phi_1&=&P_{\Pi}, \\
    \phi_2&=&P_p, \\
    \phi_3&=&P_{\alpha}, \\
    \phi_4&=&P_{\beta}, \\
    \phi_5&=&P_N, \\
    \phi_6&=& P_n, \\  
    \psi_1&=&\alpha, \\
    \psi_2&=&\beta, \\
    \chi_1&=&P_B+6m_0kA, \\
    \chi_2&=&P_A.
\end{eqnarray}
To find a set of first-class and second-class constraints, we compute the Poisson brackets between each constraint listed above, yielding
\begin{widetext}
    \begin{equation}\label{matrix_form_3}
[C_A,C_B]=
\begin{pmatrix}
0 & 0 & 0&0 &0&0&0&0&0&0\\
0 & 0 & 0&0&0&0&0&0&0&0\\
0 & 0 & 0&0&0&0&-1&0&0&0 \\
0&0&0& 0&0&0&0&-1&0&0 \\
0&0&0&0&0&0&0&0&0&0 \\
0&0&0&0&0&0&0&0&0&0 \\
0&0&1&0&0&0&0&0&0&0 \\
0&0&0&1&0&0&0&0&0&0 \\
0&0&0&0&0&0&0&0&0&6m_0k \\
0&0&0&0&0&0&0&0&-6m_0k&0\\
\end{pmatrix} \,,
\end{equation}
\end{widetext}
where de indices $A$, $B$, run from $1$ to $10$. Each number corresponds to the constraints $\phi_1$, $\phi_2$, $\phi_3$, $\phi_4$, $\phi_5$, $\phi_6$, $\psi_1$, $\psi_2$, $\chi_1$ and $\chi_2$ respectively. The rank of this $10\times10$ matrix is $6$, indicating that its nullity is $10-6=4$. Consequently, there are four independent eigenvectors with eigenvalues equal to zero. Specifically, there are four distinct combinations of the previous set of constraints that are classified as first-class constraints, given by:
\begin{eqnarray}
    C_1^{FC}&=&P_{\Pi}, \\
    C_2^{FC}&=&P_{\beta}, \\
    C_3^{FC}&=&P_N, \\
    C_4^{FC}&=&P_n.
\end{eqnarray}

Moreover, there are $10-4=6$ remaining linearly independent vectors that are associated with the set of second-class constraints. An example is displayed as follows:
\begin{eqnarray}
    C_1^{SC}&=&P_{\alpha}, \\
    C_2^{SC}&=&P_{\beta}, \\
    C_3^{SC}&=&\alpha, \\
    C_4^{SC}&=&\beta, \\
\label{C_5_second_class_Effremov}C_5^{SC}&=&P_B+6m_0kA, \\
    C_6^{SC}&=&P_A.
\end{eqnarray}
Subsequently, to derive the total Hamiltonian, the solutions for the Lagrange multipliers provided in \eqref{solution_lambda_3_Effremov} and \eqref{solution_lambda_4_Effremov} are plugged into the primary Hamiltonian as specified in \eqref{primary_hamiltonian_Effective}, resulting in
\begin{equation}\label{Total_hamiltonian_Efremov}
H_T=H'+\lambda_1C_1^{FC}+\lambda_2C^{FC}+\lambda_5C_3^{FC}+\lambda_6C_4^{FC},
\end{equation}
with
\begin{widetext}
\begin{eqnarray}
    H'&=&pC_1^{FC}+nC_3^{FC}+C_4^{SC}C_5^{SC}+2m_0kNC_2^{SC}(2NP_A-3\Pi^{1/2}C_3^{SC})+\left( \frac{p}{2\Pi}+\frac{n}{N} \right)C_2^{SC}C_4^{SC}\nonumber\\
    &+&\frac{3}{4}\frac{\Pi^{1/2}}{N}(C_3^{SC})^2+\frac{\Pi^{-1/2}}{N}(C_4^{SC})^2+\frac{1}{4m_0k}\frac{n}{N^3}C_3^{SC}C_4^{SC},
\end{eqnarray}
\end{widetext}
as the first-class Hamiltonian. Noting that the previous total Hamiltonian $H_T$ relies on constraints, we can establish that it is a gauge theory. 
\subsection{Solution to the second-class constraints}
Following the same approach outlined previously in \ref{sec_A_1}, second-class constraints will be explicitly solved in terms of canonical coordinates. The only non-trivial second-class constraint is \eqref{C_5_second_class_Effremov}, and a possible solution to it reads
\begin{eqnarray}
    A=-\frac{P_B}{6m_0k}.
\end{eqnarray}
Since only first-class constraints are present, the extended action can be expressed in the following way:
\begin{eqnarray}\label{Extended_action_First_class_effremov}
    &&S_E=\int\bigg[ P_{\Pi}\Dot{\Pi}+P_p\Dot{p}+P_B\Dot{B}+P_N \Dot{N}+P_n\Dot{n} \nonumber\\
    &&-\mu_1C_1^{FC}-\mu_2C_2^{FC}-\mu_3C_3^{FC}-\mu_4C_4^{FC} \bigg]dt, \nonumber\\
\end{eqnarray}
with
\begin{eqnarray}
    \label{C_1_solved_Effremov}C_1^{FC}&=& P_{\Pi}, \\
    \label{C_2_solved_effremov}C_2^{FC}&=& P_p, \\
    \label{C_3_solved_effremov}C_3^{FC}&=& P_N, \\
    \label{C_4_solved_effremov}C_4^{FC}&=&P_n.
\end{eqnarray}
The phase space is labeled by the following pairs of canonical coordinates: $(\Pi, P_{\Pi})$, $(p, P_p)$, $(B, P_B)$, $(N, P_N)$, and $(n, P_n)$. These pairs are accompanied by the first-class constraints \eqref{C_1_solved_Effremov}-\eqref{C_4_solved_effremov}. Using this framework, we can compute the number of degrees of freedom of this theory, resulting in the following outcome:

\begin{eqnarray}
    \mathrm{D.O.F}&=&\frac{1}{2}[10-2\times 4] \nonumber\\
    &=&1.
\end{eqnarray}
When comparing the extended actions \eqref{ext_action_3} [or \eqref{ext_action_4}] with the expression in \eqref{Extended_action_First_class_effremov}, we find that the formulation related to coupling the $4$-dimensional Palatini action with the $4$-dimensional effective theory given by \eqref{ext_action_3} [or \eqref{ext_action_4}] is equivalent to considering the $4$-dimensional effective theory on its own. It is important to note that this equivalence holds only in one of the permissible scenarios permitted for analyzing the coupling of the $4$-dimensional Palatini theory with the $4$-dimensional effective action in the Hamiltonian approach. Therefore, we point out that under the conditions of homogeneity and isotropy in the fields involved, the $4$-dimensional effective theory ``turns off'' gravity when this latter corresponds to the permissible formulation given by \eqref{ext_action_3} [or \eqref{ext_action_4}].   
 
\section{Conclusions}\label{conlcusions}
In this paper, we performed the canonical analysis of a $4$-dimensional effective action derived from the Kaluza-Klein compactification using an appropriate ansatz~\cite{efremov2014universe} coupled to the Palatini action in $4$ dimensions with or without cosmological constant, while taking into account homogeneity and isotropy in all fields involved. Due to the fact that the resulting action \eqref{action1} involves second-time derivatives and is singular as well, we employ the method of Ostrogradsky along with Dirac's algorithm~\cite{HenneauxTeitelboim+1992,dirac1964lectures} to carry out the analysis. As a result of the consistency relation of the primary constraint $\phi_8$ given by \eqref{dot_phi_8_first}, we found several scenarios to investigate. In Sec.~\ref{sec_A_1}, after solving the second-class constraints in terms of canonical coordinates, we arrive at extended actions in terms of first-class constraints only, and we realize that only three different formulations are permissible. The first corresponds to \eqref{ext_action_1}, where $\Lambda<0$ to obtain a real formulation. At the same time, the second is given by \eqref{ext_action_2}, where $\Lambda \neq 0$, and the third is represented by \eqref{ext_action_3} [or \eqref{ext_action_4}], where there is no cosmological constant in the canonical formulation. Nonetheless, it is important to emphasize here that, by a Darboux transformation~\cite{Arnold_book}, it is possible to obtain formulations \eqref{ext_action_1} or \eqref{ext_action_2} with or without a cosmological constant $\Lambda$ in terms of canonical coordinates. Moreover, we stress that the three different formulations \eqref{ext_action_1}, \eqref{ext_action_2}, and \eqref{ext_action_3} [or \eqref{ext_action_4}] are gauge theories, since the Hamiltonian corresponding to each formulation is formed only by constraints. This result is expected from the outset, as it is well known that both the $4$-dimensional Palatini action \eqref{Palatini_action} and the $4$-dimensional effective action \eqref {eff_action} are gauge theories.

In addition, the canonical analysis of the $4$-dimensional effective theory was carried out maintaining the same conditions of homogeneity and isotropy for all the fields involved. In this case, it turns out that only one formulation is permissible. After solving the set of second-class constraints, it is surprising that it coincides exactly with the formulation given in \eqref{ext_action_3} (or \eqref{ext_action_4}). Hence, we can confirm that the canonical formulation of coupling first-order gravity to $4$-dimensional effective action given by \eqref{ext_action_3} (or \eqref{ext_action_4}) is equivalent to the formulation \eqref{Extended_action_First_class_effremov}. In other words, the $4$-dimensional effective action ``turns off'' gravity only under the conditions mentioned above, causing gravity to behave as a topological theory in this valid context.

It is well known that in the $4$-dimensional Palatini Hamiltonian framework, $Q$ is a canonical coordinate with $\Pi$ as its corresponding momentum (see Appendix~\ref{canonical_analysis_Palatini}). This property changes in the canonical analysis when coupling the $4$-dimensional Palatini action to the $4$-dimensional effective theory \eqref{action1}. It turns out that after solving the second-class constraints, in the first permissible formulation \eqref{ext_action_1}, neither $Q$ nor $\Pi$ are present. In the second case \eqref{ext_action_2} $Q$ serves as a canonical coordinate, while $\Pi$ does not appear explicitly. In contrast, in the third scenario \eqref{ext_action_3} [or \eqref{ext_action_4}], the situation is reversed since $Q$ is absent from the formulation and $\Pi$ is included.

Moreover, it is worth stressing that in the first scenario \eqref{ext_action_1}, the variable related to the lapse function $N$ does not appear; whereas in the second and third scenarios given by \eqref{ext_action_2}, and \eqref{ext_action_3} [or \eqref{ext_action_4}], $N$ is a canonical coordinate whose corresponding momentum is denoted by $P_N$. Unlike the Hamiltonian analysis of the $4$-dimensional Palatini action, the lapse function $N$ is a Lagrange multiplier since there are no time derivatives on it (see Appendix~\ref{canonical_analysis_Palatini}).   

On the other hand, it would be quite interesting to investigate the canonical analysis of the $4$-dimensional Palatini action \eqref{Palatini_action} coupled to the $4$-dimensional effective action \eqref{eff_action}, preserving the space-time dependence in all fields involved, specifically by removing the assumptions of homogeneity and isotropy in those fields.

\acknowledgments
This work was supported by the SECIHTI Network Project No. 376127 {\it Sombras, lentes y ondas gravitatorias generadas por objetos compactos astrofísicos}. R.S.S. would like to thank N. Cabo-Bizet and A. Martínez-Merino for their fruitful comments. R.E. and R.H.J. are supported by SECIHTI Estancias Posdoctorales por M\'{e}xico, Modalidad 1: Estancia Posdoctoral Acad\'{e}mica and by SNII-SECIHTI. C.M. and R.S.S want to thank SNII-SECIHTI, PROINPEP-UDG, PROSNII-UDG.

\appendix
\section{No permissible cases coupling $4$-dimensional Palatini action with $4$-dimensional effective action}\label{non_permissible_Palatini}
This appendix clarifies the reasons for the cases where inconsistencies hold in the canonical analysis of the coupling of $4$-dimensional Palatini action with the $4$-dimensional effective action.

\subsection{Case $\psi=\alpha \approx0$, $\beta\neq 0$ with $Q\approx 0$ and $q\approx 0$}
In this case, we have the following constraints $\xi_1=Q$ and $\xi_2=q$, where the time evolution of each of them yields:
\begin{eqnarray}
    \Dot{\xi}_1&=& \xi_2-\frac{1}{3}N\Pi^{1/2}\Lambda \nonumber\\
    &\approx& -\frac{1}{3}N\Pi^{1/2}\Lambda, \\
    \Dot{\xi_2}&=& \lambda_2.
\end{eqnarray}
We observe that the analysis is consistent when $\Lambda=0$. Otherwise, $N$ or $\Pi$ must vanish, which contradicts the initial hypothesis. On the other hand, considering this latter result, the constraint $\chi_2$ reads,
\begin{equation}
    \chi_2=\frac{\Pi^{-1/2}}{N^2}\beta^2.
\end{equation}
Since $N\neq0$ and $\Pi\neq 0$, the constraint $\chi_2$ can be redefined as follows,
\begin{equation}
    \chi_2=\beta\approx 0.
\end{equation}
This fact contradicts our initial hypothesis given by $\beta \neq 0$. Therefore, this case is inconsistent.

\subsection{Case $\alpha\neq 0$ with $\beta\approx 0$}
Within this scenario, we recognize the constraint $\beta\approx0$, denoted as $\psi$, with $\alpha\neq0$. Considering this constraint, the set of consistency relations \eqref{dot_phi_1_first}-\eqref{dot_phi_8_first} together with $\Dot{\psi}$ reduces to,
\begin{eqnarray}
\label{dot_phi_1_case_beta}\Dot{\phi}_1&=& -6(2N\Pi^{1/2}Q+p)-6\lambda_3, \\
    \label{dot_phi_2_case_beta}\Dot{\phi}_2&=& 0, \\
    \label{dot_phi_3_case_beta}\Dot{\phi}_3&=& 3m_0k\Pi^{-1/2}\alpha NP_{\beta}-\frac{9}{8}\frac{\alpha^2}{\Pi^{1/2}N}-3N\Pi^{-1/2}Q^2 \nonumber\\
    &+&3N\Pi^{1/2}\Lambda+\frac{1}{2}\frac{\alpha P_A}{\Pi}+6\lambda_1+\frac{1}{8m_0k}\frac{\alpha}{N^2\Pi}\lambda_6, \\    \label{dot_phi_4_case_beta}\Dot{\phi}_4&=& 0, 
\end{eqnarray}
\begin{eqnarray}
\label{dot_phi_5_case_beta}\Dot{\phi}_5&=&\frac{1}{4m_0 kN^2}\lambda_6, \\
    \label{dot_phi_6_case_beta}\Dot{\phi}_6&=&-P_B-\frac{1}{2}\frac{pP_{\beta}}{\Pi}-\frac{n}{N}P_{\beta}-6m_0kA \nonumber \\
    &+&\frac{1}{4m_0k}\frac{\alpha n}{N^3}-\frac{1}{4m_0k}\frac{\alpha}{N^2}\left( \frac{\lambda_3}{2\Pi}-\frac{\lambda_7}{N} \right) \nonumber\\
    &-&\frac{1}{4m_0kN^2}\lambda_5, \\
    \label{dot_phi_7_case_beta}\Dot{\phi}_7&=& (-8m_0kNP_A+6m_0k\Pi^{1/2}\alpha)P_{\beta}-\frac{3}{4}\frac{\Pi^{1/2}}{N^2}\alpha^2 \nonumber\\
    &-&6\Pi^{1/2}Q^2+2\Pi^{3/2}\Lambda+\frac{\alpha}{N}P_A-\frac{1}{4m_0k}\frac{\alpha}{N^3}\lambda_6, \\
    \label{dot_phi_8_case_beta}\Dot{\phi}_8&=& 0\\
    \label{dot_psi_case_beta}\Dot{\psi}&=& 4m_0kN^2P_A-6m_0k\Pi^{1/2}\alpha N+\lambda_6.
\end{eqnarray}
From \eqref{dot_phi_1_case_beta}, \eqref{dot_phi_3_case_beta} and \eqref{dot_phi_5_case_beta} we find the solutions for the Lagrange multipliers $\lambda_3$, $\lambda_1$ and $\lambda_6$ given by:
\begin{eqnarray}
    \label{lambda_1_beta}\lambda_1&=&-\frac{1}{2}m_0k\Pi^{-1/2}\alpha NP_{\beta}+\frac{3}{16}\frac{\Pi^{-1/2}\alpha^2}{N}+\frac{1}{2}N\Pi^{-1/2}Q^2 \nonumber\\
    &-&\frac{1}{2}N\Pi^{1/2}\Lambda-\frac{\alpha P_A}{12\Pi}, \\
    \label{lambda_3_beta}\lambda_3&=&-(2N\Pi^{1/2}Q+p), \\
    \label{sol_lambda_6_beta}\lambda_6&=&0.
\end{eqnarray}
Considering \eqref{sol_lambda_6_beta}, the relations \eqref{dot_phi_7_case_beta} and \eqref{dot_psi_case_beta} become tertiary constraints, since there are no conditions for any Lagrange multiplier. More explicitly:
\begin{eqnarray}
    \label{chi_1_beta}\chi_1&=&(-8m_0 kNP_A+6m_0k\Pi^{1/2}\alpha)P_{\beta}-\frac{3}{4}\frac{\Pi^{1/2}}{N^2}\alpha^2 \nonumber\\
    &-&6\Pi^{1/2}Q^2+2\Pi^{3/2}\Lambda+\frac{\alpha}{N}P_A, \\
    \label{chi_2_beta}\chi_2&=&2NP_A-3\Pi^{1/2}\alpha.
\end{eqnarray}
Furthermore, $\lambda_1$ and constraint $\chi_1$ given by \eqref{chi_1_beta} can be rewritten as follows:
\begin{eqnarray}
\label{chi_1_simplyfied_beta}\chi_1&=&-2m_0k(\chi_2+2NP_A)\phi_6-\frac{\alpha}{4N^2}\chi_2+\frac{\alpha}{N}P_A \nonumber\\
    &-&6\Pi^{1/2}Q^2+2\Pi^{3/2}\Lambda \nonumber\\
    &\approx&\frac{\alpha}{N}P_A-6\Pi^{1/2}Q^2+2\Pi^{3/2}\Lambda,
\end{eqnarray}
where constraints \eqref{phi_6_Palatini_Effremov} and \eqref{chi_2_beta} were employed. Furthermore, $\lambda_1$ given in \eqref{lambda_1_beta} can be rewritten as follows,
\begin{equation}
\label{lambda_1_beta_simplyfied}\lambda_1\approx -\frac{5}{48}\frac{\Pi^{-1/2}\alpha^2}{N}-\frac{1}{3}N\Pi^{1/2}\Lambda.
\end{equation}
Subsequently, the application of the consistency relations to the constraints $\chi_1$ and $\chi_2$ given by \eqref{chi_1_simplyfied_beta} and \eqref{chi_2_beta} reads,
\begin{eqnarray}
    \label{dot_chi_1_beta}\Dot{\chi}_1&=& \frac{P_A}{N}\lambda_5-\frac{\alpha P_A}{N^2}(n+\lambda_7) \nonumber\\
    &+&6N\Pi^{1/2}Q(\Pi^{-1/2}Q^2-\Pi^{1/2}\Lambda) \nonumber\\
    &-&12\Pi^{1/2}Q(q-\frac{5}{48}\frac{\Pi^{-1/2}\alpha^2}{N}-\frac{1}{3}N\Pi^{1/2}\Lambda), \\
\label{dot_chi_2_beta}\Dot{\chi}_2&=&2P_A(n+\lambda_7)+3\alpha NQ-3\Pi^{1/2}\lambda_5, 
\end{eqnarray}
where the solutions for $\lambda_1$ and $\lambda_3$ provided by \eqref{lambda_1_beta_simplyfied} and \eqref{lambda_3_beta} were utilized.

On the other hand, plugging back $\lambda_3$ given by \eqref{lambda_3_beta} into \eqref{dot_phi_6_case_beta} yields,
\begin{eqnarray}\label{dot_phi_6_beta_simplyfied}
    \Dot{\phi}_6&=&-P_B-\frac{1}{2}\frac{pP_{\beta}}{\Pi}-\frac{n}{N}P_{\beta}-6m_0kA+\frac{1}{4m_0k}\frac{\alpha n}{N^{3}} \nonumber\\
    &+&\frac{1}{4m_0k}\frac{\alpha}{2\Pi N^2}(2N\Pi^{1/2}Q+p) \nonumber\\
    &+&\frac{1}{4m_0k N^2}(\frac{\alpha}{N}\lambda_7-\lambda_5).
\end{eqnarray}
Observing \eqref{dot_chi_1_beta}, \eqref{dot_chi_2_beta}, and \eqref{dot_phi_6_beta_simplyfied}, we identify three equations with two unknowns $\lambda_5$ and $\lambda_7$. This scenario prevents the analysis from yielding a consistent solution.

\subsection{Case $\psi_1=\alpha\approx 0$, $\psi_2=\beta\approx0$ with $Q\approx 0$ and $q\neq 0$}
We have the following constraint $\xi=Q\approx0$. The evolution in time of this latter reads:
\begin{equation}
    \Dot{\xi}=q-\frac{1}{2}N\Pi^{1/2}\Lambda.
\end{equation}
In the absence of any condition of the Lagrange multiplier, an additional constraint is introduced, herein designated as $\Omega$,
\begin{equation}
    \Omega=q-\frac{1}{2}N\Pi^{1/2}\Lambda.
\end{equation}
At this stage of the analysis, we find $\Lambda\neq 0$. Otherwise, $q$ vanishes. On the other hand, from the relation of the constraint $\xi_2$ given by
\eqref{chi_2_second_case}, we find that
\begin{eqnarray}
    \chi_2&=&3Q^2-\Pi\Lambda \nonumber \\
    &\approx&-\Pi \Lambda.
\end{eqnarray}
The latter implies $\Lambda=0$, which contradicts the result given by $\Lambda\neq 0$. Therefore, this case is inconsistent. 
\\

\section{No permissible cases for $4$-dimensional effective action}\label{non_permissible_Effremov}
This appendix explains the cases where inconsistencies hold in the canonical analysis of the $4$-dimensional effective action.

\subsection{Case $\alpha\approx0$ with $\beta\neq 0$}
In this case, we recognize $\psi=\alpha$ as the secondary constraint. Applying the consistency relation, we get,
\begin{equation}
    \Dot{\psi}=\lambda_3,
\end{equation}
which implies,
\begin{equation}\label{lambda_3_Effremov_alpha}
    \lambda_3=0.
\end{equation}
Thus, bearing in mind $\psi=\alpha$ as a constraint and \eqref{lambda_3_Effremov_alpha}, the set of consistency relations for each constraint given by \eqref{Dot_phi_1_Effremov}-\eqref{dot_phi_6_Effective} together with \eqref{lambda_3_Effremov_alpha} collapses to,  
\begin{eqnarray}
\label{dot_phi_1_effremov_alpha_simplyfied}\Dot{\phi}_1&=&\frac{1}{2}\frac{\Pi^{-3/2}}{N}\beta^2, \\
    \Dot{\phi}_2&=& 0, \\
    \Dot{\phi}_3&=&-\frac{1}{4m_0k N^2}\left[ \frac{\beta}{N}(n-\lambda_5)+\frac{\beta}{2}\lambda_1-\lambda_4 \right], \\
    \Dot{\phi}_4&=&-P_B-6m_0kA-\frac{2\Pi^{-1/2}\beta}{N}, \\
    \Dot{\phi}_5&=&\frac{\Pi^{-1/2}}{N^2}\beta^2, \\
    \Dot{\phi}_6&=&0.
\end{eqnarray}
Since $N\neq0$ and $\Pi\neq 0$, we find that \eqref{dot_phi_1_effremov_alpha_simplyfied} yields $\beta\approx 0$, which contradicts our initial assumption.  

\subsection{Case $\alpha\neq 0$ with $\beta\approx 0$}
Regarding $\psi:=\beta$ as the secondary constraint, the set of consistency relations related to each primary constraint \eqref{Dot_phi_1_Effremov}-\eqref{dot_phi_6_Effective} is rewritten as follows: 
\begin{eqnarray}
    \label{dot_phi_1_effremov_beta}\Dot{\phi}_1&=& 3m_0k\Pi^{-1/2}\alpha NP_{\beta}-\frac{9}{8}\frac{\alpha^2}{\Pi^{1/2}N}+\frac{1}{2}\frac{\alpha P_A}{\Pi} \nonumber \\
    &+& \frac{1}{8m_0k}\frac{\alpha}{N^2\Pi}\lambda_4, \\
    \label{dot_phi_2_effremov_beta}\Dot{\phi}_2&=&0, \\
    \label{dot_phi_3_effremov_beta}\Dot{\phi}_3&=&\frac{1}{4m_0kN^2}\lambda_4, \\
    \label{dot_phi_4_effremov_beta}\Dot{\phi}_4&=&-P_B-\frac{1}{2}\frac{pP_{\beta}}{\Pi}-\frac{n}{N}P_{\beta}-6m_0kA+\frac{1}{4m_0k}\frac{\alpha n}{N^3} \nonumber\\
    &-&\frac{1}{4m_0k}\frac{\alpha}{N^2}\left( \frac{\lambda_1}{2\Pi}-\frac{\lambda_5}{N} \right)-\frac{1}{4m_0kN^2}\lambda_3, \\
    \label{dot_phi_5_effremov_beta}\Dot{\phi}_5&=& (-8m_0kNP_A)P_{\beta}-\frac{3}{4}\frac{\Pi^{1/2}}{N^2}\alpha^2+\frac{\alpha}{N}P_A, \\
    \label{dot_phi_6_effremov_beta}\Dot{\phi}_6&=&0.
\end{eqnarray}
From \eqref{dot_phi_3_effremov_beta}, we can fix $\lambda_4$ as follows:
\begin{equation}\label{lambda_4_beta_effremov}
    \lambda_4=0.
\end{equation}
Moreover, using \eqref{lambda_4_beta_effremov}, the consistency relation for the secondary constraint $\psi=\beta$ collapses to:
\begin{equation}
    \Dot{\psi}=-2m_0kN(-2NP_A+3\Pi^{1/2}\alpha).
\end{equation}
Since there is no condition for any Lagrange multiplier, we have the following tertiary constraint:
\begin{equation}\label{chi_beta_effremov}
    \chi=-2NP_A+3\Pi^{1/2}\alpha.
\end{equation}
Thus, from \eqref{phi_4_effremmov}, \eqref{lambda_4_beta_effremov}, and \eqref{chi_beta_effremov}, the previous consistency relations \eqref{dot_phi_1_effremov_beta} and \eqref{dot_phi_5_effremov_beta} acquire the following form,
\begin{eqnarray}
    \label{dot_phi_1_effremov_beta_sym}\Dot{\phi}_1&=&\frac{3}{8}\frac{\alpha^2}{\Pi^{1/2}N}, \\
    \label{dot_phi_5_effremov_beta_sym}\Dot{\phi}_5&=&-\frac{3}{4}\frac{\Pi^{1/2}}{N^2}\alpha^2.
\end{eqnarray}
Given that $N\neq 0$ and $\Pi\neq 0$, and based on the recent results, \eqref{dot_phi_1_effremov_beta_sym} or \eqref{dot_phi_5_effremov_beta_sym}, we conclude that $\alpha\approx0$ must hold. However, this conclusion contradicts our initial assumption. 

\section{Canonical analysis of Palatini action at homogeneous and isotropic approximation}\label{canonical_analysis_Palatini}
In this section, we carry out the canonical analysis of the Palatini action with cosmological constant, bearing in mind that the fields involved are homogeneous and isotropic (see~\cite{PhysRevD.101.084003,PhysRevD.101.024042} for a complete canonical analysis in a manifestly Lorentz covariant fashion without second-class constraints). After this imposition, the Palatini action collapses to \eqref{Palatini_action_cosmological}. Here, the pair of canonical variables are $(\Pi,Q)$, which satisfy the canonical Poisson bracket, 
\begin{equation}
    \lbrace \Pi(t),Q(t) \rbrace=\frac{1}{6}.
\end{equation}
In this case, the lapse function $N$ plays the role of a Lagrange multiplier, as it does not contain any time derivatives. This fact implies that the unique constraint is given by the well-known scalar constraint given by
\begin{equation}
    \tilde{\mathcal{H}}:=6\Pi^{1/2}Q^2-2\Pi^{3/2}\Lambda.
\end{equation}
Furthermore, $\Pi$ is the canonical momentum of $Q$, and these variables do not have a canonical momentum associated with each other, as happened above with the coupling of the $4$-dimensional Palatini action with the $4$-dimensional effective theory. It is worth mentioning that the variable $Q$ is related to the Ashtekar connection $A$ as follows (see for instance~\cite{PhysRevD.101.084003,PhysRevD.101.024042} to observe the relation between the Lorentz vector $Q_{a}{}^{i}$ and the connection Asthekar variables $A_{a}{}^{i}$ when there are no assumptions of homogeneity and isotropy in the fields):
\begin{equation}
    Q=-\frac{n^0}{\gamma}A,
\end{equation}
with $n^0$ as the temporal component of \eqref{Lorentz_vector_n}, and $\gamma$ as the Immirzi parameter~\cite{Giorgio_Immirzi_1997}. 

%%%%%%%%%%%%%%%%%%%%%%%%%%%%%	
\bibliographystyle{apsrev4-1}
	
\bibliography{coupling_gravity}

\end{document}